\documentclass[twocolumn]{aastex62}

\defcitealias{Simard2011BulgeDiskDecompCatalog}{S11}

\begin{document}

\title{\MakeUppercase{Improved GALEX UV Photometry for 700,000 SDSS galaxies}}

\author{Chandler Osborne}
\affil{Department of Astronomy, Indiana University, Bloomington, IN, 47408}

\author[0000-0003-2342-7501]{Samir Salim}
\affil{Department of Astronomy, Indiana University, Bloomington, IN, 47408}

\author[0000-0003-0946-6176]{Médéric~Boquien}
\affiliation{Instituto de Alta Investigación, Universidad de Tarapacá, Casilla 7D, Arica, Chile}

\author[0000-0001-5414-5131]{Mark Dickinson}
\affil{NSF’s National Optical-Infrared Astronomy Research Laboratory, 950 N. Cherry Ave., Tucson, AZ 85719, USA}

\author{Stéphane Arnouts}
\affil{Aix Marseille University, CNRS, CNES, LAM, Marseille, France}

\begin{abstract}

The Galaxy Evolution Explorer (GALEX) satellite performed the first and only large-area UV survey, which in tandem with the Sloan Digital Sky Survey (SDSS) has facilitated modeling of the spectral energy distributions (SEDs) of low-redshift galaxies and the determination of various galaxy properties, in particular the star formation rate. However, the relatively crude angular resolution of GALEX (5\arcsec{}) made its images susceptible to blending of sources, resulting in potentially biased far-UV (FUV) and near-UV (NUV) pipeline photometry. To remedy this issue and take advantage of model-fit photometry, we use the EMphot software to obtain forced GALEX photometry for $\sim$700,000 SDSS galaxies at z$<0.3$. Positional priors of target galaxies and potentially contaminating neighbors were taken from SDSS. New photometry is based on the best-fitting of three model profiles: optical-like, exponential and flat. New photometry mitigates blending present in the original pipeline catalogs, which affected $16\%$ of galaxies at a level of $>0.2$ mag and $2\%$ at a level of $>1$ mag. Pipeline NUV magnitudes are severely affected ($\gtrsim1$ mag) when the neighbor is brighter than the target galaxy and within 10\arcsec{}, or when the neighbor is fainter and within $\sim$3\arcsec{} of the target. New photometry fixes edge-of-detector bias, which affected pipeline photometry by up to 0.1 mag in NUV. We present catalogs with new photometry for GALEX observations of different depths, corresponding to the all-sky imaging survey (AIS), medium imaging survey (MIS) and deep imaging survey (DIS). Catalogs feature combined magnitudes for multiple detections of the same galaxy in a survey. 

\end{abstract}

\section{Introduction}

To understand the evolution of galaxies, the current star formation rate (SFR) and stellar mass must be accurately constrained.  Both properties can be estimated by fitting model spectra to observed spectral energy distributions, a method referred to as the SED fitting \citep{Conroy2013SEDFittingReview}.  For optimal constraints on the SFR, the rest-frame UV ($\sim 100 - 300$ nm) should be included in the SED fitting as this is where the emission of the short-lived (O and B) stars peaks.  

UV imaging is now widely available from the Galaxy Evolution Explorer (GALEX) satellite, which imaged $77\%$ of the sky in at least one UV band.  GALEX was active from July 2003 to February 2012 and was equipped with two filters, one sensitive to the far-UV (FUV; $\sim 1500$\r{A}) and the other sensitive in the near-UV (NUV; $\sim 2300$\r{A}) with FWHM resolutions of 4.2\arcsec{} and 5.3\arcsec{}, respectively \citep{Morrissey2007GALEX}.  

The GALEX photometry produced from UV imaging by the GALEX science team is archived at Mikulski Archive for Space Telescopes (MAST).  We will be referring to the photometry in the data release GR7 as the ``pipeline'' photometry.  The GALEX pipeline photometry is based on source identification and segmentation of the UV images with SExtractor \citep{Bertin&Arnouts1996SExtractor}.  While photometry for a variety of circular aperture sizes is available, the best option for galaxies is the MAG AUTO, which adopts a Kron-like elliptical aperture based on the overall two-dimensional UV profile, as determined by SExtractor \citep{Bertin&Arnouts1996SExtractor}.  Source identification and photometry is performed independently for FUV and NUV images and a band-merged catalog of sources is produced which includes detections from both bands as a single entry.  The merged catalog also includes measures of the flux in one band at the position of the source in the other band.  

GALEX photometry is often used in conjunction with the photometry from the optical Sloan Digital Sky Survey (SDSS), which imaged the sky in a suite of optical filters (\textit{ugriz}) at a typical resolution (in \textit{r}) of $\sim 1.32$\arcsec{}, a factor of 4 (3) smaller than GALEX NUV (FUV).  The SDSS photometric pipeline employs several  methods to produce the optical photometry.  In addition to aperture fluxes, the SDSS pipeline also includes model-based photometry.  An exponential disk (S\'ersic index $= 1$) and de Vaucouleurs profile (S\'ersic index $= 4$) are separately fit to each object to determine a flux.  The photometry, as well as the best-fit parameters, such as the size for each type of model are included in the SDSS data release.  The quantity reported by SDSS pipeline as modelMag refers to the magnitude from the profile (exponential or de Vaucouleurs) that was found to be more likely (a better fit).  Model photometry is in general less sensitive to blending than the aperture photometry because the neighbors are fit as well (if detected as separate objects), or, if not detected, still the pixels with contaminating flux will be down-weighted by forcing a model profile. 

Because of the large difference in resolution between GALEX and SDSS and the differences in the type of photometry that is available for galaxies (aperture vs. model), sources that are resolved in SDSS images may be blended in the GALEX images.  As a result, GALEX pipeline magnitudes may be systematically biased with respect to SDSS magnitudes, which directly impact the reliability of SED fitting for which the goal is to reproduce the observed color using models.  Furthermore, GALEX NUV photometry is affected by edge of detector effects, where objects close to an image boundary have overestimated magnitudes by 0.1 mag \citep{Salim2016GSWLC}.  

The overarching goal of this work is to improve the SED modeling for SDSS galaxies by first improving the UV photometry.  More specifically, it will be used to improve the SED fitting results from the GALEX-SDSS-WISE Legacy Catalog (GSWLC), an SED fitting catalog of $\sim700,000$ galaxies at $z < 0.3$ \citep{Salim2016GSWLC, Salim2018DustAttCurves}.  Uncertainties and biases in the UV photometry propagate and affect all aspects of the SED modeling, biasing estimation of the SFR and other properties.  SDSS galaxies are used as a baseline for our understanding of the more distant universe, so it is essential to ensure the accuracy of GALEX photometry and the properties derived from it. 

Robust deblending can be achieved via forced photometry, where object positions and shapes from a higher resolution survey are used to extract photometry from a lower resolution survey.  Forced photometry has the additional benefit of boosting the signal-to-noise of fluxes of very faint objects, helping to improve the detection rates.

Several forced photometry codes are available publicly, including EMphot \citep{Conseil2011EMphot}, TPHOT \citep{Merlin2016TPHOT}, and the Tractor \citep{Lang2016TheTractor}.  Forced photometry for wide-area surveys has already been used to great effect, notably in the unWISE data release which performed forced photometry on WISE near/mid-IR images (resolution of 5\arcsec{} to 12\arcsec{}) using the Tractor software with SDSS priors \citep{Lang2016unWISEforcedphotometry, Lang2016TheTractor}.  Our goal is to use a similar approach to create a catalog of forced photometry for low-redshift ($z < 0.3$) galaxies in GALEX.  To this end, we make use of the EMphot software \citep{Guillaume2006GALEXEMphotDeblendingDeepFields, Llebaria2008SPIEGALEXEMphotOpticalPriors, Vibert2009SPIEEMphot, Conseil2011EMphot}, which was developed specifically for forced photometry on GALEX images using optical priors. EMphot is already setup to work with GALEX pipeline imaging products. EMphot has previously been used primarily to extract GALEX photometry of deep fields where galaxies are often unresolved in GALEX images \citep[e.g.,][]{Ilbert2009COSMOSphotoZ, Salim2009MidIRDeepFields, Moustakas2013PRIMUSConstraintsOnQuenchingMergingAndSMF, EMphotMoutardCFHTLS2016}, but not for more nearby galaxies.  

To perform forced photometry we will require positions of target galaxies and their neighbors from SDSS, as well as some assumption about the UV light profile.  The UV emission in general tends to be more stochastic or `clumpy' than the optical \citep{Marcum2001UVAtlas&Morphology}.  Furthermore, galaxies which are bulge-dominated in the optical may be disk-dominated in the UV, since star formation occurs primarily in disks.  Some disks have flatter profiles in the UV compared to even the exponential profiles \citep{GildePaz2007GALEXUVAtlas}.  We will account for the diversity in UV profiles by allowing for exponential and flat models in addition to optical-like models that contain both the disk and the bulge.  

The paper is organized as follows.  In Section \ref{Section:Data&SS} we describe the data used and the sample of galaxies for which we obtain forced GALEX photometry.  Section \ref{Section:Approach&Methods} outlines our general approach and methods, Section \ref{Section:Results} presents our evaluation of the EMphot photometry, and we conclude in Section \ref{Section:Conclusion}.  



\section{Data \& Sample Selection}
\label{Section:Data&SS}

The target sample of this work are galaxies included in the GALEX-SDSS-WISE Legacy Catalog \footnote{GSWLC is available at:  \url{https://salims.pages.iu.edu/gswlc/}} \citep[GSWLC;][]{Salim2016GSWLC}.  GSWLC provides physical property estimates for $\sim 700,000$ SDSS spectroscopic galaxies in the range $0.01 < z < 0.3$ with \textit{r} $< 18$ mag. A galaxy is included in GSWLC regardless of there being a detection in UV or mid-IR, as long as it falls within the GALEX imaging footprint. In this section, we describe the datasets used, selection process for our sample, and the structural parameters used to model the photometric profiles our galaxies.    

\subsection{GALEX}

GALEX imaged $77\%$ of the entire sky and $90\%$ of the area included in SDSS DR10. Individual GALEX observations consist of circular `tiles' 1.2 degrees in diameter.  Individual exposures (`visits') ranged in duration from $\sim 100$ seconds up to $\sim 30$ minutes \citep{Morrissey2007GALEX}.  Tiles with longer exposure times are the result of co-addition of smaller individual exposures.  The largest sky coverage (27,000 sq. deg) comes from the shallow All-sky Imaging Survey (AIS), with typical exposure times of $\sim 100$s.  The Medium Imaging Survey (MIS) provides medium-depth ($\sim 1500$s) images (over 5,000 sq. deg), whereas the Deep Imaging Survey (DIS) has the smallest sky coverage (365 sq. deg), provides the deepest images obtained by co-adding multiple visits ($\sim 30,000$s).  GALEX tile images have a pixel scale of 1.5 \arcsec/pixel.  Originally, different survey names were tied to different imaging campaigns, and some observations were taken as part of campaigns with different names, such as the guest investigator program. In order to take advantage of all GALEX imaging regardless of the original campaign, following \citet{Salim2016GSWLC} we define AIS (GSWLC-A) as all tiles with NUV exposure time (in seconds) $t_{NUV} \leq 650$, MIS (GSWLC-M) as $650 < t_{NUV} < 4000$, and DIS (GSWLC-D) as $t_{NUV} \geq 4000$.  

GSWLC-A, -M and -D contain 640,659 galaxies, 361,328 galaxies and 48,401 galaxies, respectively. Each of these catalogs is available in one of two versions (1 or 2), both which contain the same objects, the difference being that for GSWLC-2 the dust luminosity as inferred by WISE photometry is directly included as a constraint in the SED fitting.  Also available is GSWLC-X (659,229 galaxies), a master catalog containing all unique galaxies and parameters associated with the deepest of either GSWLC-A, -M, or -D.  Since every galaxy from GSWLC-A, -M and -D must be present in GSWLC-X,  we use the latter to define the list of objects on which to perform the photometry, but we obtain photometry for each object in GSWLC-A, M and D lists from the tiles of respective depth. 

In summary, we perform forced photometry on 9,781 tiles corresponding to GSWLC-A, 5,398 tiles corresponding to GSWLC-M, and 621 corresponding to GSWLC-D. If an object from a given catalog appears in multiple tiles of the same type, the principal photometry comes from the tile of the longest exposure time in FUV (or longest exposure time in NUV, if FUV is not available).  






\subsection{SDSS}

Our target sample is the GSWLC-X2 catalog (described in the previous section) which includes 659,229 objects that we refer to as targets.  To account for potential blending of targets with neighboring objects, we also perform photometry on optical sources that lie within 20\arcsec{} of a target, have $u < 22$, and are flagged as primary objects by SDSS; no other cuts are applied to select the neighbors. Neighbors may consist of other GSWLC objects, foreground stars, artifacts, or galaxies outside the redshift and brightness cuts of GSWLC.  We use SDSS Casjobs to get optical (\textit{r}-band) profiles and photometry for the targets plus 321,156 neighbors (not including neighbors that are also GSWLC objects), for 980,385 total objects (targets $+$ neighbors) for which new GALEX photometry will be obtained.  The separation cut of 20\arcsec{} limits our selection to probable blends, while the magnitude cut was determined by visual inspection of isolated sources in MIS images in order to exclude objects that would not be detectable in the UV anyway.  The actual $u$ detection limit is survey dependent, but selecting a uniform cut helps reduce potential biases and makes the processing easier.  Our comparison of the photometry obtained from UV images of different depth (Section \ref{Section:UVsurveyComp}) reveals no issues with the photometry in either AIS or DIS related to the $u$ magnitude cut for neighbors.  

For each object we extract from SDSS DR16 the best-fit parameters for both profiles will be used only in cases when two-component profiles from \citet{Simard2011BulgeDiskDecompCatalog} are not available. 

We use the \citet[][hereafter S11]{Simard2011BulgeDiskDecompCatalog} catalog of bulge+disk decompositions of $r < 18$ mag SDSS galaxies as the basis of our model photometry.  Most ($\sim 96\%$) of our target sample is included in \citetalias{Simard2011BulgeDiskDecompCatalog}.  The remaining $4\%$ are excluded because \citetalias{Simard2011BulgeDiskDecompCatalog} uses SDSS DR7 and excludes $r < 14$ galaxies, whereas GSWLC uses SDSS DR10 and does not exclude bright galaxies.  For the one type of model light profile that is intended to mimic the optical profile (optical-like; see Section \ref{Section:UVprofiles}), we make use of \citetalias{Simard2011BulgeDiskDecompCatalog}'s `canonical' \textit{r}-band fits, which fix the S\'ersic index \citep{Sersic1963Profile}, of the bulge at $n_{b} = 4$ and the S\'ersic index of the disk at $n_{d} = 1$.  For a small number of galaxies with anomalous \citetalias{Simard2011BulgeDiskDecompCatalog} fits, i.e., where the half-light radius of the bulge exceeds that of the disk ($\sim4\%$ of targets), we instead make use of their single-component fits for which total S\'ersic index was allowed to vary freely. For the $4\%$ of target galaxies that are not in \citetalias{Simard2011BulgeDiskDecompCatalog} we use exponential ($n_{d} = 1$) or de Vaucouleurs ($n_{d} = 4$) single-component profiles from SDSS official catalog  \citep{Ahumada2020SDSSDR16datarelease}.

\section{Approach and Methods}
\label{Section:Approach&Methods}

\subsection{Motivation}

\begin{figure}
    \centering
    \includegraphics[scale=0.7]{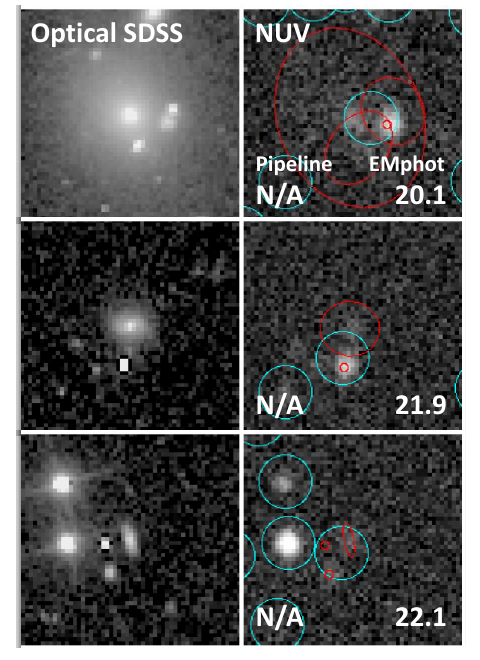}
    \caption{Examples of target galaxies for which no pipeline detection is available due to either blending or incompleteness.  The optical SDSS \textit{r}-band image (rescaled to GALEX pixel scale) is shown on the left and the GALEX NUV image is shown on the right.  Optical priors are outlined in red while GALEX pipeline detections are marked with cyan circles (10\arcsec{} radius).  Images are $\sim1.3$\arcmin{} on each side.  The target galaxy in the first row (a large elliptical) likely lacks an individual detection in the GALEX pipeline catalog due to a blending with the object to the right.  The target in the second row may be missed in the GALEX pipeline catalog because of the brighter detection nearby.  The target in the third row may suffer from a combination of blending and other bright sources being nearby.}
    \label{panel3_eximages}
\end{figure}

\begin{figure*}
    \centering
    \includegraphics[scale=0.7]{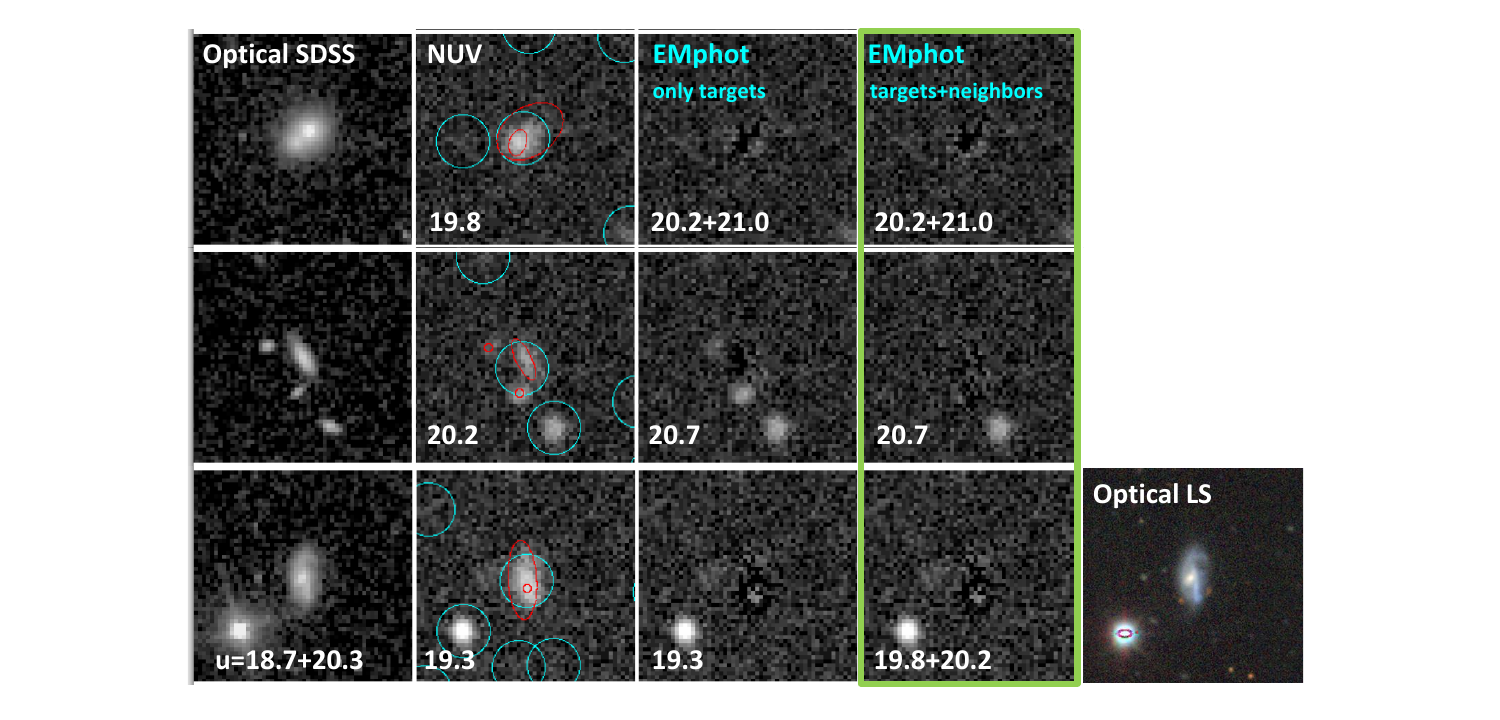}
    \caption{A set of example images showing instances where multiple SDSS objects are blended in the GALEX pipeline photometry.  Each row is a unique object; the optical SDSS image is shown in the first column, the original GALEX NUV is shown in the second, and best EMphot residuals are shown for runs that do not model the neighbors (i.e., contain only GSWLC targets; third column) and with neighbors (fourth column).  Red ellipses correspond to optical prior positions and shapes, while cyan circles (10\arcsec{} radius) are pipeline detections.  Images are $\sim1.3$\arcmin{} on each side.  In the first row, the two optical objects are both GSWLC targets and their EMphot fluxes add up accurately to the blended pipeline photometry value.  In the second row, the pipeline photometry is a blend of the GSWLC target plus its two neighbors; despite the blending, EMphot is able to recover a more appropriate flux even when neighbors are not included in the fitting.  The third row shows a galaxy which has an embedded starburst region identified as a separate object by SDSS (the \textit{u}-band magnitudes for both objects are shown; a color composite image from LS \citep{Dey2019LegacySurvey} better shows the starburst region on the right).  This starburst region is not included in the SDSS photometry, though the GALEX pipeline and the EMphot run without neighbors do not exclude it.  Inclusion of the neighboring objects in the EMphot fitting is necessary to obtain photometry that is consistent with SDSS.}
    \label{panel2_eximages}
\end{figure*}

We first illustrate the improvements in photometry afforded by the use of optical source positions and model shapes (referred to hereafter as optical priors).  In Figure \ref{panel3_eximages} we show examples of target galaxies which lack a NUV pipeline detection within 5\arcsec{} of the SDSS position, even when the UV source is distinguishable by eye.  The reason for this usually appears to be the presence of another bright source in the vicinity.  By using optical priors together with EMphot, we are able to recover fluxes for targets missed by the pipeline or `lost' due to blending.

Next, we demonstrate the effectiveness of EMphot in deblending of sources.  In Figure \ref{panel2_eximages} we show examples of target galaxies which are blended with other objects in the GALEX pipeline, and show the performance of EMphot in two cases: one where only target galaxies are modeled, and another where both targets and neighbors are modeled.  When the pipeline photometry is clearly a blend of multiple sources, we see that the EMphot fluxes add up accurately to the blended pipeline magnitude; this is clearly shown in the first and third rows of Figure \ref{panel2_eximages}.  Even if neighbors are not included in the EMphot run, EMphot is often able to account for blending due to the constraints obtained by forcing a profile; this is evident in the second row of Figure \ref{panel2_eximages}.  In other cases, including the neighbor in the modeling is necessary to deblend the flux (row 3 of Figure \ref{panel2_eximages}).  

\subsection{EMphot Description}

EMphot version 2.0 is an IDL program designed to compute UV photometry from GALEX images using object positions and light profiles (or images) from a higher-resolution optical band as priors.  EMphot was originally developed to study deep GALEX tiles with overlap in the CFHTLS deep fields \citep{EMphotMoutardCFHTLS2016}, i.e., faint galaxies with small angular diameters.  EMphot works by following a series of steps.  First, the input list of prior positions and optionally shapes and/or images are used to produce model UV profiles.  EMphot then applies astrometry corrections and degrades the input image (if present) to GALEX resolution and convolves it with the GALEX PSF.  After determining an initial UV flux, an expectation-maximization algorithm is employed to fit the amplitudes of model profiles to a background-subtracted image.  EMphot does not allow for negative fluxes. Output includes the flux catalog, simulated UV image, and residual or difference image (subtraction of simulated UV image from the real UV image).  Photometric error is determined using the residual image, i.e., the difference between the fitted prior profile and real UV image, using a quadratic sum over the residuals of each pixel associated with a given object. 

We note that EMphot uses only the GALEX PSF on the input images. An SDSS to GALEX kernel would normally be needed when taking SDSS images and degrading them to make GALEX models. However, this is not relevant for this study because we provide EMphot with synthetic images which have zero PSF. While our model parameters do come from SDSS which has a non-zero PSF, the SDSS PSF is used in the determination of the model parameters, so there is no need for EMphot to use anything more than the GALEX PSF in the image convolution.

Aside from the specification of the PSF profile which differs between NUV and FUV, there is no functional difference in using EMphot on NUV or FUV images.  For a complete technical description of EMphot, see \citet{Conseil2011EMphot}.  

EMphot can be run in one of three different modes.  First is `dirac' mode, which uses only the positions of the optical priors and assumes each object to be unresolved in the UV; i.e., the UV profile being fit is simply the UV PSF.  Dirac mode is not appropriate for our purposes since our target galaxies are largely resolved.  Second is `shape' mode, which uses positions and shape information (size, half-light radius, axis ratio, and position angle) to create a more realistic but still simplified model profile, an exponential disk.  The `shape' mode is essentially equivalent to SDSS model photometry using an exponential profile.  A pure exponential profile will not always or even often be a good assumption for a UV light profile, so we do not use this mode either.  Third is `stamp' mode, which combines positions and shape information with the optical image to create a detailed, non-parametric light profile.  Given that the UV light profiles are often very different from the optical ones, we do not use this mode either.  Our goal is to perform model photometry, but with a range of profiles in addition to the exponential one. To achieve this we use stamp mode, but supply it with the synthetic images consisting of different model profiles of our own choosing (see Section \ref{Section:UVprofiles}).  

To run EMphot, one needs to supply (for a specific GALEX tile) the intensity, sky background, and relative response images, along with the pipeline source catalog. For stamp mode, a prior catalog of object positions and shapes is also required as well as the image (in our case a synthetic one) containing the priors. 

\begin{figure}
    \centering
    \includegraphics[scale=1]{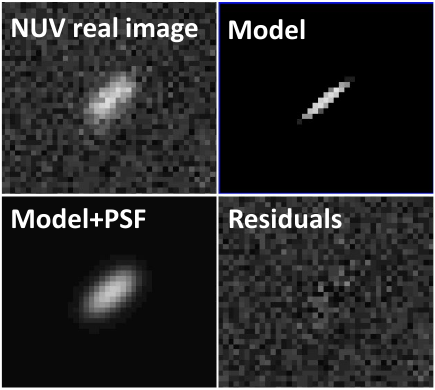}
    \caption{Comparison of the real NUV image (top left), best model input image (top right), PSF-convolved model image (bottom left), and residual image (bottom right) for a galaxy in our sample.  This particular galaxy is best fit by a flat profile (Sersic $n = 0.25$).  The image cutouts are 1\arcmin{} on each side.}
    \label{panel4_process_ex}
\end{figure}

Figure \ref{panel4_process_ex} shows the real, input model, PSF-convolved model, and residual images for one target galaxy which is best fit by a flat profile with Sersic $n = 0.25$.  EMphot convolves the input model with the GALEX PSF, then fits the resulting model+PSF profile to the real image.  The photometric error is determined by the residual image, which is the difference between the real and model+PSF fit image.  For more details on the synthetic image generation, see Section \ref{Section:UVprofiles}.  

\subsection{Model Light Profiles}
\label{Section:UVprofiles}

\begin{figure*}
    \centering
    \includegraphics[scale=0.7]{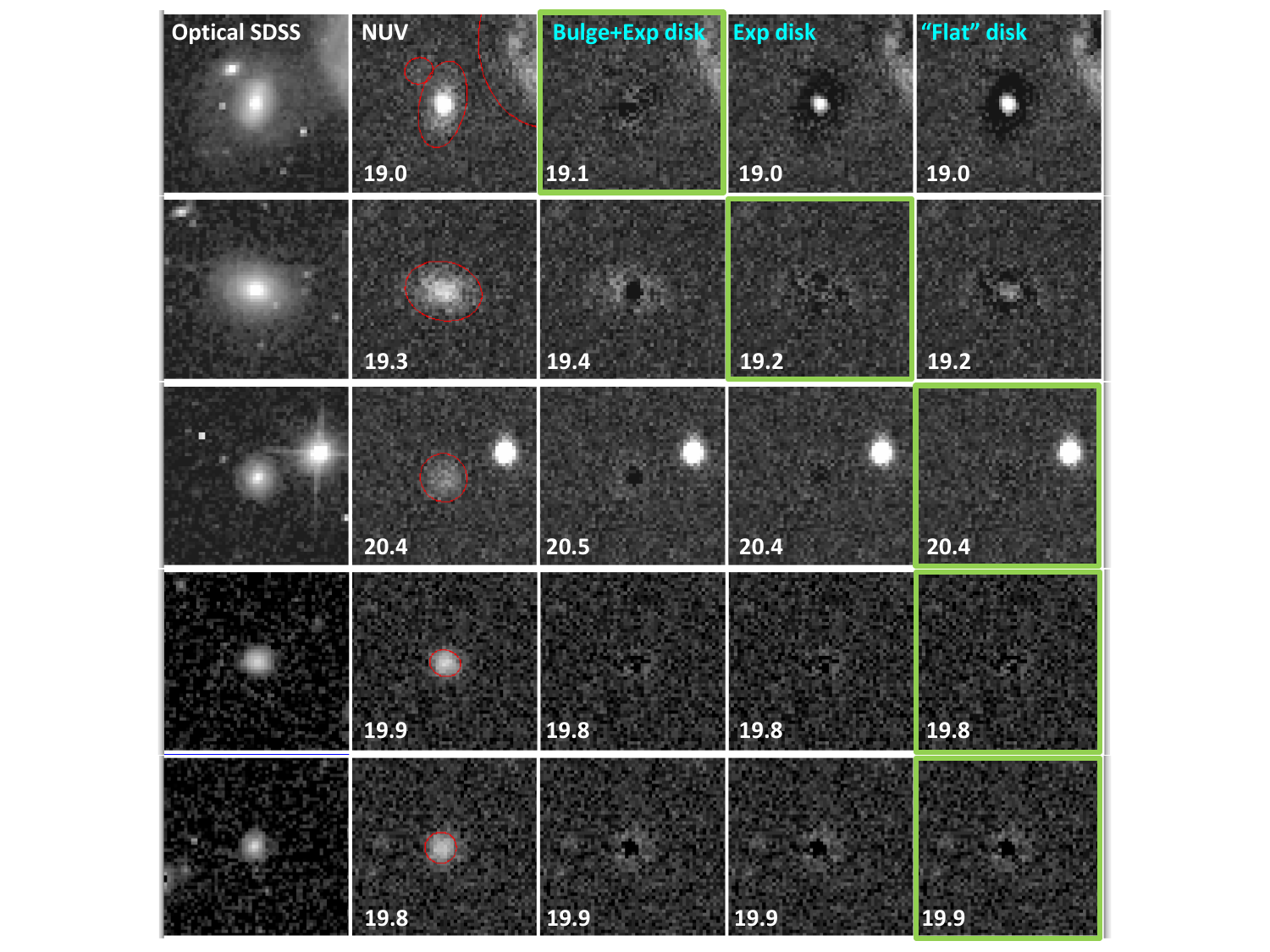}
    \caption{Example galaxy images showing the effectiveness of different types of UV profiles to obtain photometry.  Images are $\sim1.3$\arcmin{} on each side.  Each row corresponds to a different galaxy.  The first column shows the SDSS \textit{r}-band image resampled to the GALEX pixel scale.  The second column shows the NUV image with the pipeline magnitude listed in the lower-left corner, and red outlines corresponding to the optical prior positions and shapes.  The next three columns show the residual images and magnitudes for optical-like (bulge+disk), exponential, and flat profiles.  Panels with the best-fitting EMphot profiles are marked with green borders.  The galaxy in the fourth row has similar residuals with all three profiles.  Although the galaxy in the bottom row is not well modeled by any profile due to the extended UV emission, the best EMphot magnitude is still close to the pipeline magnitude (only 0.04 mag difference).}
    \label{panel1_eximages}
\end{figure*}

In this work we wish to produce an equivalent of SDSS model photometry, but for the UV.  Out of the box, EMphot can perform model photometry only with exponential profiles.  An exponential profile is too shallow for light of galaxies with prominent bulges, even in the UV, whereas some late-type galaxies have UV light profiles even shallower than the exponential \citep{GildePaz2007GALEXUVAtlas}.  To facilitate EMphot model photometry with profiles other than exponential, we utilize its `stamp' mode, but instead of using an actual image we feed it with synthetic images featuring one of the three different types of model profiles at the time: optical-like, exponential, and flat.  In other words, we perform photometry of target galaxies three times, each time assuming a different profile.  The `best' profile is the one which minimizes the error in relative flux (i.e., minimizes the error in magnitude), which is to say that it has the smallest residuals.  More precisely, EMphot determines the flux error for an object using a quadratic sum of the residuals, weighted by the optical profile convolved with the GALEX PSF, over all pixels associated with that object \citep[see][]{Conseil2011EMphot}.

Here we describe our three profile treatments:\newline

%

\noindent \textbf{Optical-like profiles:}  To model target galaxies with optical-like profiles, we adopt 2-component S\'ersic (bulge, $n = 4$ plus disk, $n = 1$) profiles from \citetalias{Simard2011BulgeDiskDecompCatalog}.  Profiles are truncated at 3.5 times the half-light radius.  We find the optical-like profile to be the best choice for fainter (in UV) galaxies since they are more often the early types whose UV profiles are similar to optical ones (being mostly from older stars).  We find that optical-like profiles best fit the UV emission for $43\%$ of target galaxies.  \newline

\noindent \textbf{Exponential disk profiles:}  We also model the UV emission as an exponential disk, i.e., a single S\'ersic profile with $n = 1$.  For targets contained in \citetalias{Simard2011BulgeDiskDecompCatalog} we adopt the half-light radius of the disk component.  In rare cases ($\sim4\%$ of targets), the \citetalias{Simard2011BulgeDiskDecompCatalog} half-light radius for the disk component is anomalously small.  In such cases, we instead use the total galaxy half-light radius from the \citetalias{Simard2011BulgeDiskDecompCatalog} single-component S\'ersic (free $n$) fit.  The profiles are again truncated at 3.5 times the half-light radius.  We find that this type of profile best fits the UV emission for $28\%$ of target galaxies.  \newline


\noindent \textbf{Flat profiles:}  This implementation uses a single S\'ersic model with $n = 0.25$.  Such profiles are essentially flat out to the effective radius and then drop quickly \citep{Ciambur2016Profiler}, following qualitatively the UV profiles of many nearby galaxies in \citet{GildePaz2007GALEXUVAtlas} as well as the Type V profiles of \citet{Binggeli1991SurfaceBrightnessProfiles} characteristic of some dwarf elliptical galaxies.  The adopted half-light radius and size are the same as for the pure exponential profile.  We find that this type of profile best fits the UV emission for $29\%$ of target galaxies. \newline


Model profiles are centered on the RA and DEC positions of the optical priors.  We use Astropy \citep{Astropy2013, Astropy2018} to handle most of the image processing and to construct the profiles.  Synthetic images used by EMphot are created with the same pixel scale (1.5\arcsec{}) and size as the real GALEX images.  

Neighbors of our target galaxies (SDSS objects within 20\arcsec{} of target galaxies) are typically small and faint, making the model profile available from official SDSS processing often very unreliable. We thus model them as unresolved sources, except in rare cases when they are bright enough to be included in \citetalias{Simard2011BulgeDiskDecompCatalog}, in which case we model them as we model the target galaxies.  In most cases the neighbors are
either foreground stars or fragments of larger target
galaxies.  Treating these fragments as separate objects is justified because they were treated the same way in SDSS, with which we want to have consistent magnitudes.  In rare cases a target may overlap with a very large foreground galaxy with angular extent greater than our neighbor distance limit of 20\arcsec{}. Using the S4G \citep{Sheth2010S4GInitialRelease, Watkins2022S4GExtended} and LVL \citep{Dale2009SpitzerLVLSurvey} catalogs of nearby galaxies, we visually inspected the GALEX NUV images of all targets lying beyond 20\arcsec{} of an S4G/LVL source but still within 3 times the half-light radius of the best-fitting \textit{r}-band SDSS pipeline model for that S4G/LVL galaxy. We found 23 targets which have some degree of blending in the NUV images. These 23 targets are flagged in the final catalog as having possible contamination in their UV fluxes, and are excluded from all subsequent analysis in this work. 

We have also considered including two additional types of profile.  The first is a modification of the 2-component optical-like profile in which we make the bulge two times (0.75 mag) fainter than it is in the optical image; this is intermediate between the optical-like and exponential profiles and serves as a `proxy' for profiles based on bluer bands like \textit{u} or \textit{g}.  The other profile we considered is a single-component S\'ersic $n=0.5$ profile intermediate between the exponential and flat profiles.  We ran photometry with these additional profiles on one test tile. For all objects in this tile, we have performed careful inspection of the residuals and evaluation of the photometry in cases that were better fit by one of these two profiles than by our nominal three.  We found that the fraction of objects that are best fit by these profiles is relatively small compared to the total number of objects ($\sim 10\%$ for half-bulge and $\sim 15\%$ for $n=0.5$), plus they have only marginally smaller photometry errors (differences of order 0.01 mag or less). In all cases, one of our nominal three profiles was as good as these two in terms of the visual appearance of the residuals (i.e., there was no difference in the quality of the fit upon visual inspection), and there were no systematics in the photometry.  We conclude that the scheme with three profiles is sufficiently robust, with the benefit of keeping the processing times reasonable, at $\sim 4$ minutes per tile ($\sim 263$ hours total for AIS + MIS + DIS). 

In Figure \ref{panel1_eximages}, we show several examples of galaxies fit with the different UV profiles.  Galaxies with UV-bright bulges are fit well with the optical-like bulge+disk profile, but not with the pure exponential or flat disk profiles; an example is shown in the first row of Figure \ref{panel1_eximages}.  Conversely, some galaxies have UV profiles that are less centrally concentrated than the optical, and so are better fit by profiles which exclude the bulge; for example, the galaxy in the second row is best described by the exponential disk profile, whereas the galaxy in the third row favors the flat disk profile.  The galaxy in the fourth row is fit well by all three profiles with little difference between the residuals.  In rare cases where no profile works particularly well, such as the galaxy in the bottom row, the EMphot magnitude is still reliable as it agrees reasonably well with the pipeline magnitude.  While having the appropriate profile improves the resulting photometry, the magnitudes based on different profiles are quite similar, with a median difference between the brightest and the faintest magnitude for the same object of only 0.08 mag.

\begin{figure}
    \centering
    \includegraphics[scale=0.4]{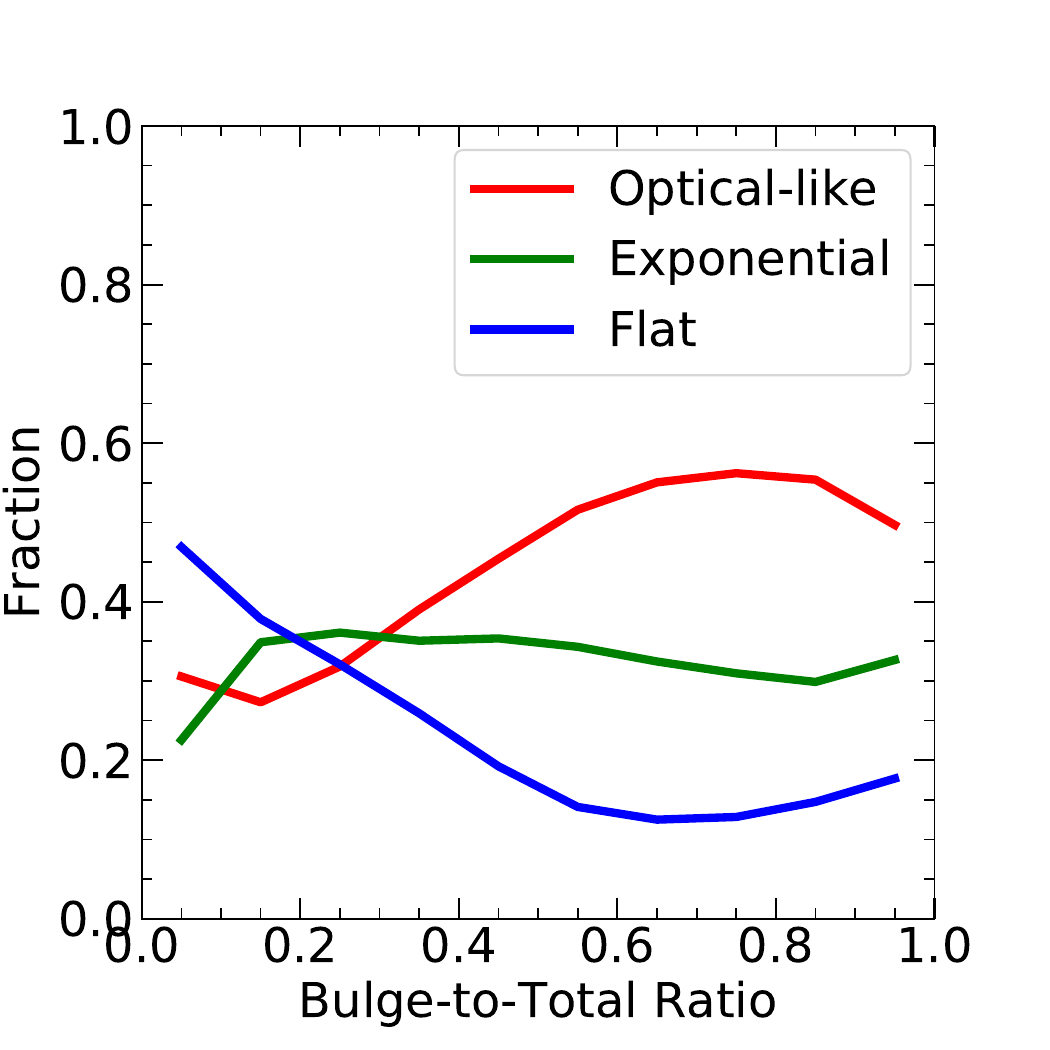}
    \caption{Fraction of galaxies with different best-fitting NUV profiles as a function of optical bulge-to-total (B/T) ratio.  Flat profiles are more common at low B/T whereas optical-like profiles are more common at high B/T.}
    \label{bestruns_versus_BtoT}
\end{figure}

Figure \ref{bestruns_versus_BtoT} shows the fraction of different best-fitting NUV profiles as a function of the optical bulge-to-total ratio (B/T) as determined by \citet{Simard2011BulgeDiskDecompCatalog}.  Flat profiles are more common at lower B/T, increasing from $\sim20\%$ at B/T$\gtrsim0.5$ to $\sim50\%$ at B/T $\sim 0$.  Flat profiles were found by \citet{GildePaz2007GALEXUVAtlas} to be associated with true late-type spirals, which is consistent with the trend we see. Optical-like profiles are more common at higher B/T, increasing from $\sim30\%$ at B/T$\sim0$ to $\sim50-60\%$ at B/T$>0.5$. We expect the optical-like profile to be more common at high B/T since it is the only profile to include a bulge. Exponential profiles are relatively constant in fraction ($\sim35\%$) with B/T, though below B/T$\lesssim0.1$ there is a decrease in their fraction to $\sim20\%$, possibly because as B/T approaches zero the exponential profile becomes difficult to distinguish from the optical-like profile.

Moving on to the FUV, the FUV light distribution may differ from that of the NUV in particular cases, but in general it is closely correlated \citep[see Figures 4j and 4k in ][]{GildePaz2007GALEXUVAtlas}.  We find from inspecting the objects in our test tile that the best NUV profile (that is, the NUV profile with the lowest error) and the best FUV profile are the same for $70\%$ of the relatively bright (NUV $< 22$ mag) objects.  Furthermore, for the purposes of galaxy SED modeling it is preferable for FUV photometry to be performed consistently with the NUV, in a similar way that SDSS modelMag magnitudes in each band are based on the best-fitting profile in \textit{r} band. Thus, when running EMphot on the FUV image for a given tile, we use only one prior image wherein for each galaxy the best-fitting profile from the NUV is adopted.  This saves a significant amount of time, as instead of performing three separate EMphot runs like for NUV, we need only perform one.  

Before finalizing our approach, in a single test tile (selected to be a typical tile with a range of galaxy types) we visually inspected galaxies for which EMphot and pipeline magnitude differed by more than $\sim0.35$ mag (EMphot usually being fainter) and found that in almost all cases the EMphot photometry was more reasonable than the pipeline one (e.g., there was obvious blending).

\section{Evaluation of UV model photometry}
\label{Section:Results}

In this section we present comparisons between the model-based EMphot photometry and the original GALEX pipeline photometry. We consider the difference between EM and pipeline magnitudes in various regimes to assess the effects of EMphot deblending and identify regimes where EMphot offers improvements.  The MIS strikes a balance between sky coverage and depth, and furthermore has a narrower range in exposure times (more uniform depth for individual tiles) compared to either AIS or DIS.  For this reason, we largely focus on MIS (i.e, galaxies in GSWLC-M listing) to assess the improvements and possible biases in the model photometry, but also include a comparison of both MIS and AIS to the deeper DIS. 

\begin{figure}
    \centering
    \includegraphics[scale=0.55]{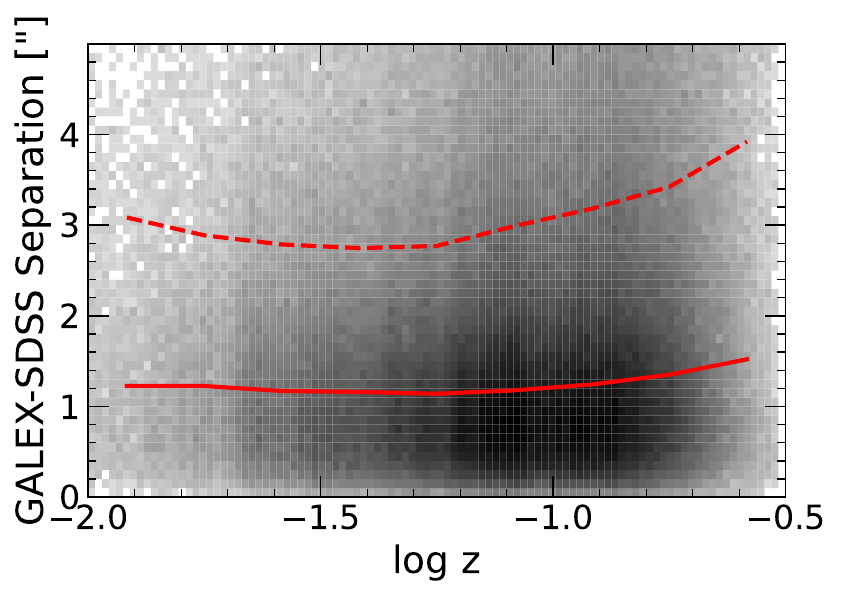}
    \caption{Separation between the GALEX pipeline and SDSS source positions for all AIS galaxies. The median separation at different redshifts is shown as a solid red line, while the 90th percentile of separations is shown as a dashed red line. The median separation is $\sim 1.2$\arcsec{} at all redshifts.}
    \label{distance_versus_z}
\end{figure}

We use the SDSS-GALEX matching from GSWLC where the GALEX match to an SDSS object is the brightest (in NUV) object within 5\arcsec{} of the SDSS position \citep{Salim2016GSWLC}. The $1\sigma$ positional uncertainty (for isolated SDSS sources) is $\sim 0.2$\arcsec{} for SDSS and $\sim1.3$\arcsec{} for GALEX \citep{Salim2016GSWLC}. Based on the positional uncertainty, the fraction of galaxies with SDSS-GALEX separations larger than 5\arcsec{} is $0.3\%$. Figure \ref{distance_versus_z} shows the GALEX-SDSS separation versus redshift for all AIS objects. The median separation is $\sim 1.2$\arcsec{} at all redshifts (solid red line in Figure \ref{distance_versus_z}). Furthermore, $90\%$ of galaxies have separations within $\sim3$\arcsec{} at $z\sim0.01$ and within $\sim 4$\arcsec{} at $z \sim0.3$ (dashed red line in Figure \ref{distance_versus_z}). We therefore find the matching to be highly robust.

Aside from the random error which is determined directly from the EMphot residuals, there may be a systematic error arising from our assumptions about the UV profiles.  We quantify this systematic error by considering the profile mismatch error, i.e., the difference in magnitude between the best and second best fitting profiles.  For MIS galaxies, the mismatch error is $20\%$ of the random error on average, which amounts to only $2\%$ of the total error (i.e., the sum of the two errors in quadrature).  Therefore, we find that the uncertainties of the EMphot photometry are dominated by the random errors rather than systematics due to our assumptions on the UV profiles.  

\subsection{EMphot detection rates}

\begin{figure}
    \centering
    \includegraphics[scale=0.6]{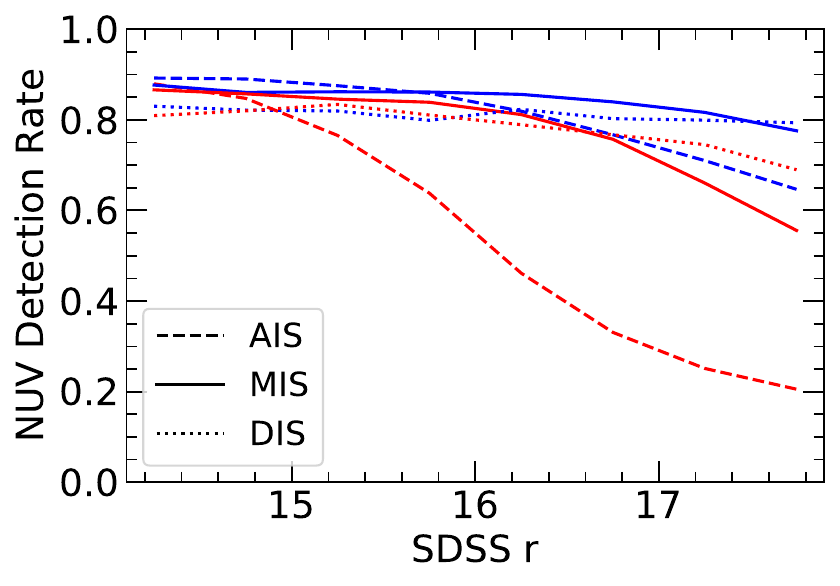}
    \caption{NUV detection rates of target galaxies based on new (EMphot) photometry as a function of SDSS \textit{r}-band magnitude. Threshold to be considered a detection is $3\sigma$ (i.e., signal-to-noise ratio $> 3$). AIS galaxies are shown with dashed lines, MIS with solid lines, and DIS with dotted lines.  Galaxies are divided into blue ($u-g < 1.6$; blue lines) and red ($u-g > 1.6$; red lines).  Detection rates are generally lower for fainter magnitudes and redder colors.}
    \label{EM_completeness_GSW}
\end{figure}

Figure \ref{EM_completeness_GSW} shows EMphot NUV detection rates ($3\sigma$) for target galaxies.  We evaluate completeness for target galaxies split by optical color into blue ($u-g < 1.6$) and red ($u-g > 1.6$).  In this section we consider a $3\sigma$ detection to be a galaxy with a signal-to-noise ratio of at least 3 in the EMphot NUV photometry.  Altogether, the $3\sigma$ NUV (FUV) detection rate for MIS is $78\%$ ($55\%$).  Detection rates are lower among fainter or redder galaxies. Red AIS galaxies have the lowest overall detection rates by a large margin.  Furthermore, we find that EMphot is able to provide some estimate of the photometry (regardless of the SNR) for $79\%$ of galaxies which originally lacked a detection in the pipeline. We note that on occasion EMphot will fail to process some of the galaxies that are close to the edge of a tile. The fraction of targets which have pipeline NUV photometry but which lack EMphot NUV photometry (regardless of SNR) is $9\%$ for AIS, $11\%$ for MIS, and $15\%$ for DIS.

\begin{figure*}
  \centering
  \gridline{\fig{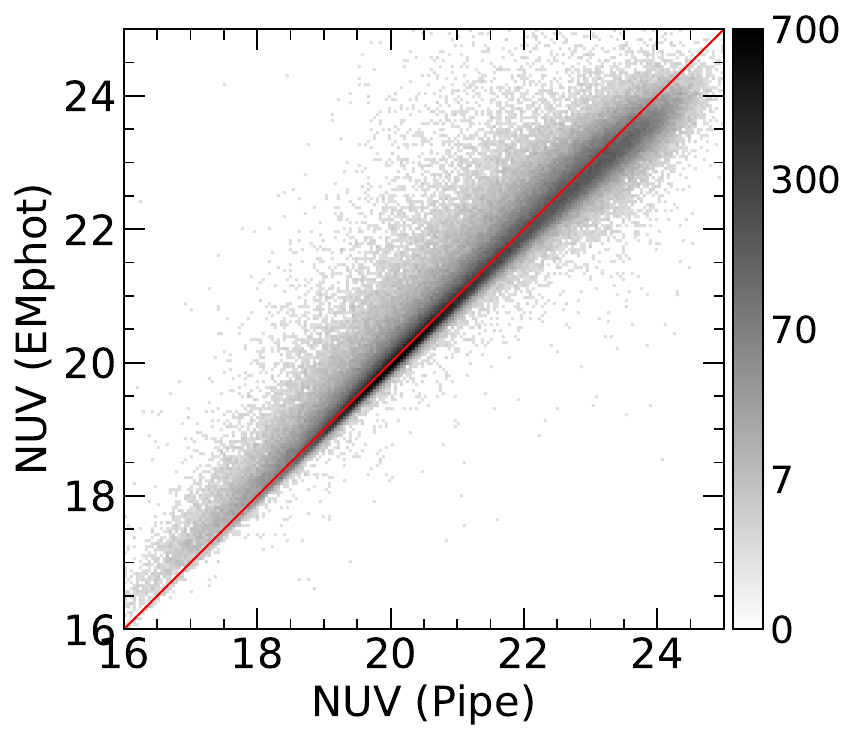}{0.3\textwidth}{}
          \fig{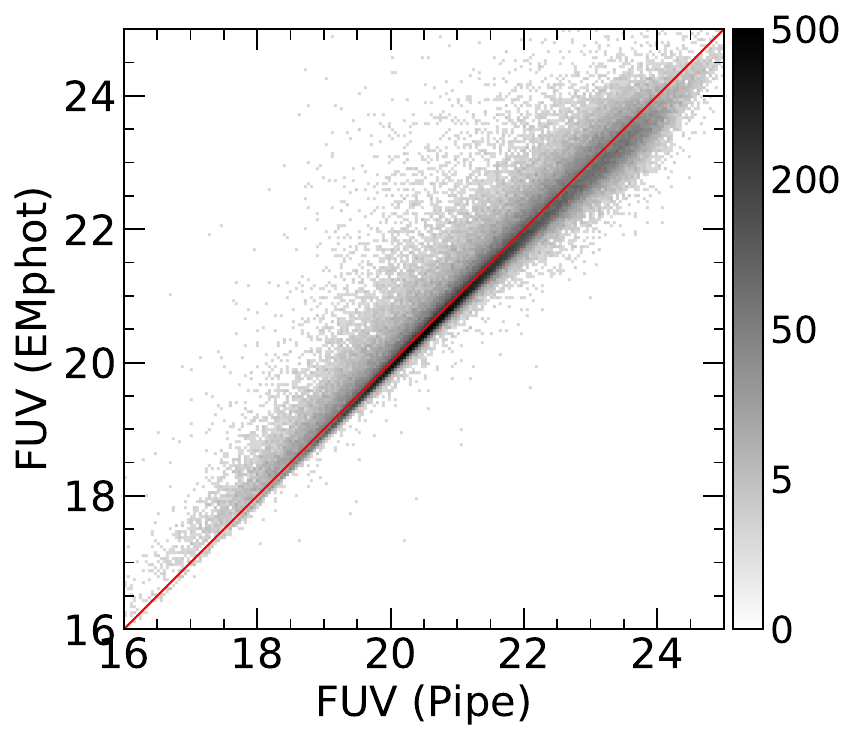}{0.3\textwidth}{}
          \fig{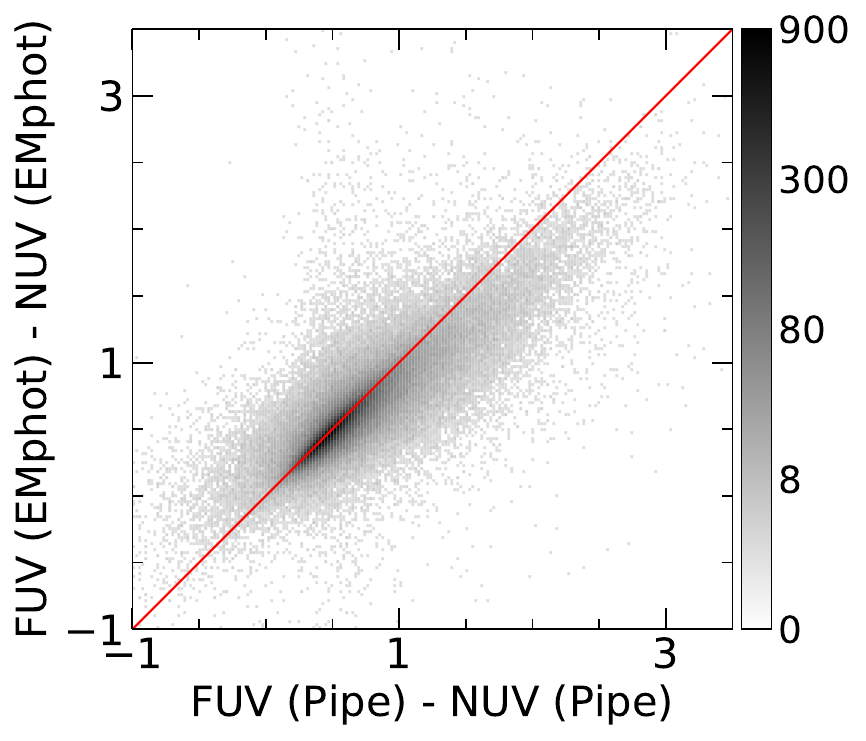}{0.3\textwidth}{}
          }
  \vspace*{-\baselineskip}
  \caption{Comparisons between new EMphot photometry to the GALEX pipeline photometry for target galaxies in the medium-deep imaging survey (MIS).  We show comparisons for NUV and FUV magnitudes as well as UV color.  The 1:1 line is shown in solid red.  Most galaxies lie on or close to the 1:1 line, but there is preferential scatter towards fainter EM magnitudes, due to the blending present in the pipeline photometry.  UV colors are generally preserved except among the very red galaxies for which EMphot derives bluer colors.  Each panel contains $\sim 320,000$ objects and we use a grayscale density ``scatterplot''. The shade of grey scales as $n^{0.3}$ in order to highlight regions with fewer objects. The same plotting scheme is used in all subsequent figures of this type.}
  \label{EM_pipe_1to1_comps}
\end{figure*}

\subsection{Consistency with pipeline photometry}

\begin{figure}
    \centering
    \includegraphics[scale=0.5]{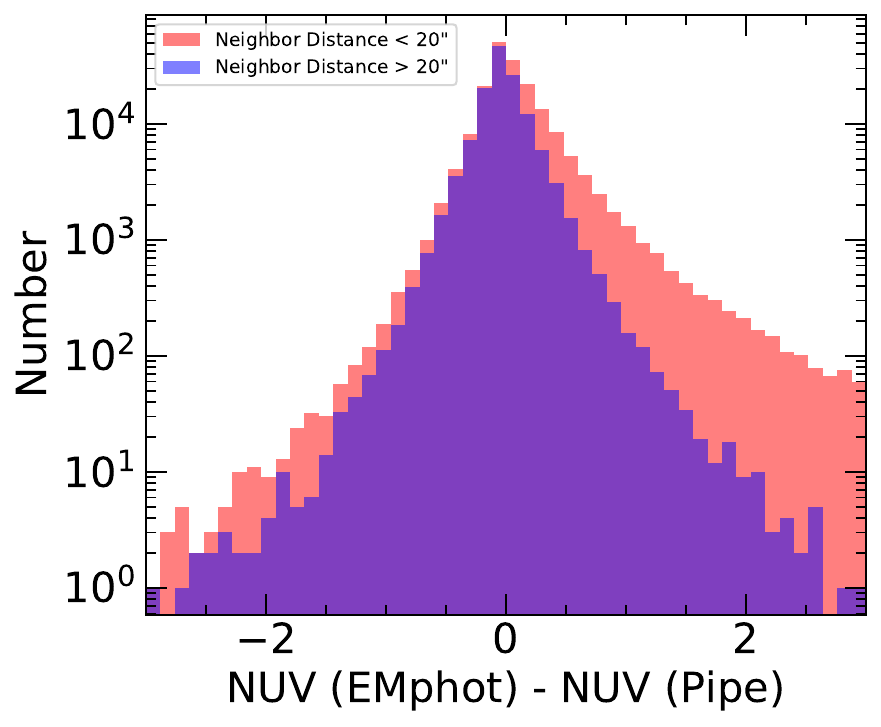}
    \caption{Number distribution in NUV (EM) - NUV (Pipe) for galaxies with (red) and without (blue) neighbors ($u<22$ SDSS objects within $20$\arcsec{}).  Galaxies with neighbors show a stronger tail to high values of NUV (EM) - NUV (Pipe), which is attributable to blending contamination.  Note that the presence of a neighbor within $20$\arcsec{} does not necessarily lead to blending (see Section \ref{Section::EffectsOfBlending}).}
    \label{deltaNUV_histogram}
\end{figure}

In Figure \ref{EM_pipe_1to1_comps} we present 1:1 comparisons of EMphot and pipeline UV magnitudes and UV colors of MIS galaxies.  Most galaxies follow the 1:1 line, showing the general consistency between the two methods, i.e., the absence of zero point offsets.  In the NUV comparison (left panel), there is preferential scatter in the sense that EMphot magnitudes are fainter, which is the result of the ability of EMphot to deal with the blending.  This is supported by Figure \ref{deltaNUV_histogram}, which shows that galaxies with neighbors (neighbor distance $< 20$\arcsec{}) have a stronger tail towards high values of NUV (EM) - NUV (Pipe) compared to galaxies without neighbors.  At the faint end (NUV (Pipe) $\gtrsim 22$), we see a bending away from the 1:1 line such that EMphot magnitudes are systematically brighter than pipeline magnitudes by up to 0.5 mag at NUV = 24.  This may be attributable to improved estimation of the background by EMphot, a conclusion supported by the results of Section \ref{Section:UVsurveyComp}, where we show agreement between magnitudes of galaxies detected in both the MIS and DIS tiles.  The bulk FUV trends (middle panel) are altogether similar to those in the NUV. 

The right panel of Figure \ref{EM_pipe_1to1_comps} shows the comparison of UV color (FUV$-$NUV) from EMphot versus that from the pipeline.  Most galaxies lie close to the 1:1 line.  However, for a small number of very red galaxies there is a noticeable trend where EMphot predicts bluer colors compared to pipeline by up to 0.5 mag.  The difference may also be due to EMphot's improved background estimation compared to the pipeline, a conclusion supported by the comparison to photometry from DIS (Section \ref{Section:UVsurveyComp}).  

GALEX pipeline catalog includes not only independent FUV detections (which we use throughout this paper) but also FUV photometry measured at the positions and with aperture sizes of NUV detections ({\fontfamily{cmtt}\selectfont FUV{\_}NCAT}).  We have verified that the overall trends are the same regardless of whether independent or NUV-based FUV photometry is adopted as a comparison for EMphot. 

\begin{figure*}
    \centering
    \gridline{\fig{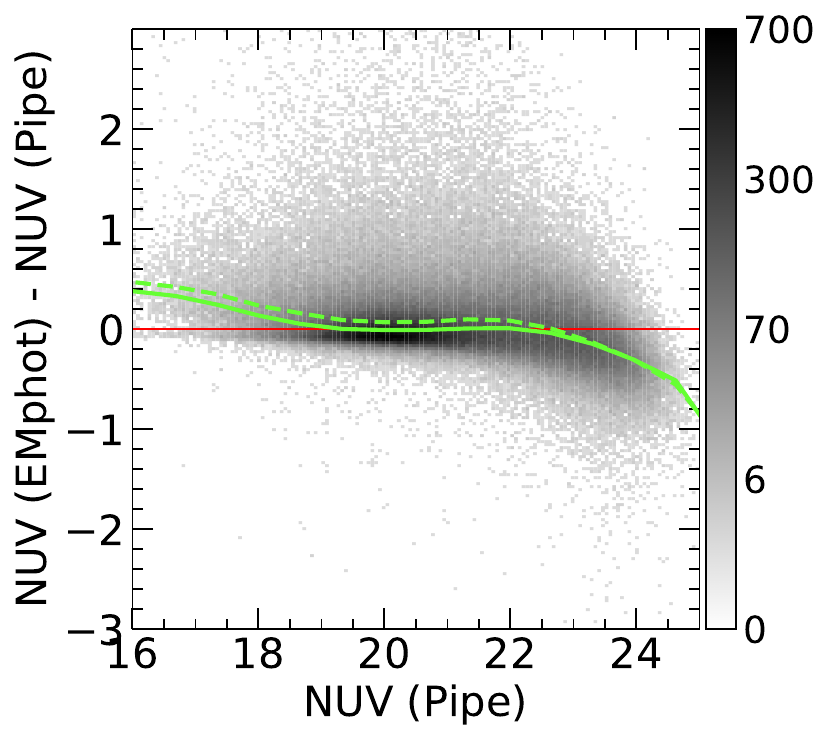}{0.3\textwidth}{}
          \fig{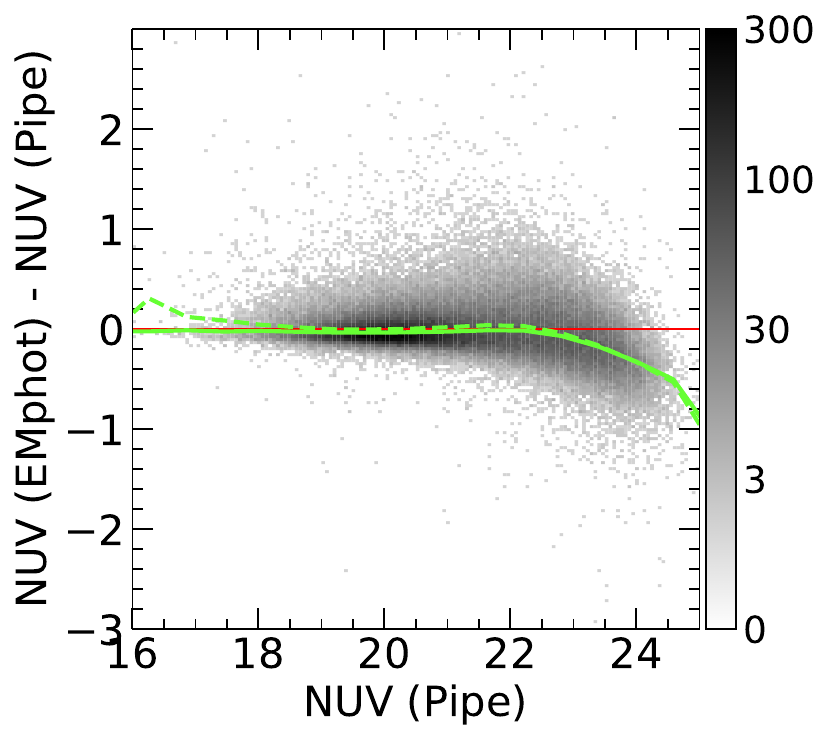}{0.3\textwidth}{}
          \fig{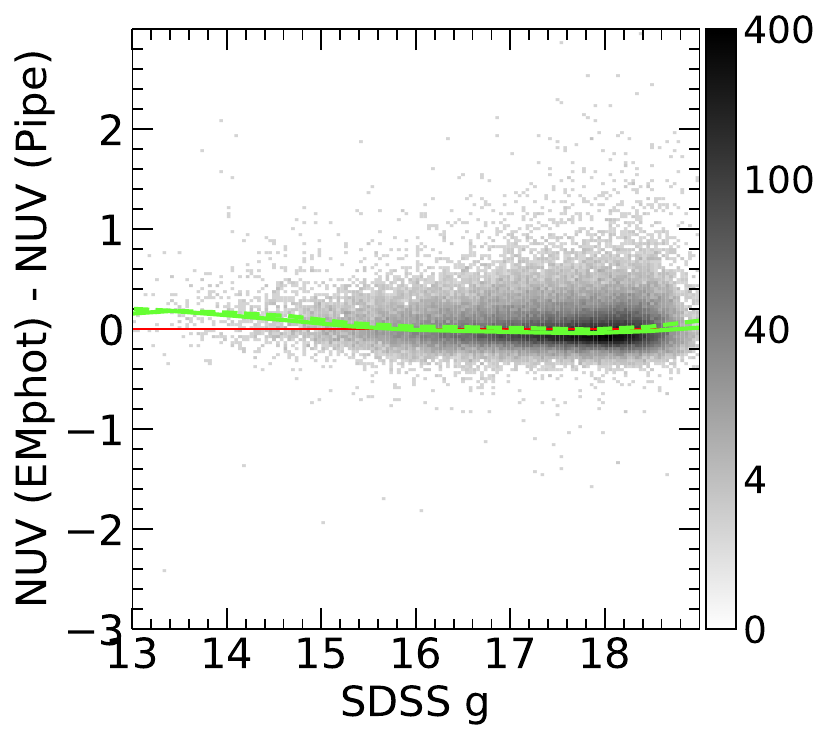}{0.3\textwidth}{}
          }
    \vspace*{-\baselineskip}
    \caption{Difference in EMphot and pipeline NUV magnitudes versus NUV pipeline magnitude (left, middle) and SDSS \textit{g} magnitude (right).  The left panel includes all of the MIS sample.  The middle and right panels show a subset of MIS galaxies with no nearby neighbors and thus no known blending.  The right panel further excludes galaxies with NUV (Pipe) $> 22$.  The solid (dashed) green lines represent the running median (mean).}
    \label{delta_vs_pipeNUV}
\end{figure*}

In order to further characterize the comparison, next we assess the difference in EMphot and pipeline magnitudes as a function of the pipeline NUV and SDSS \textit{g}-band magnitudes in Figure \ref{delta_vs_pipeNUV}.  Focusing first on the left panel of Figure \ref{delta_vs_pipeNUV}, we see that EMphot magnitudes are systematically dimmer among bright (NUV (Pipe) $< 19$) galaxies, which may be due to blending.  To test this, in the middle panel of Figure \ref{delta_vs_pipeNUV} we exclude galaxies which have neighbor ($u < 22$ object within 20\arcsec{}).  The median offset for NUV (Pipe) $< 19$ is now reduced to zero, whereas the mean offset is greatly reduced but still non-zero, which may be because the model photometry will exclude faint contaminants even if not explicitly recognized as neighbors (see Figure \ref{panel2_eximages} and the associated discussion). 

In the right panel of Figure \ref{delta_vs_pipeNUV}, we keep our focus on objects without neighbors, but pipeline NUV from the middle panel is replaced with SDSS \textit{g} and furthermore a NUV (Pipe) $< 22$ cut is applied to exclude faint targets affected by differences in background subtraction already discussed. EMphot magnitudes are systematically dimmer by up to 0.2 mag among the bright ($g < 16$) galaxies even though we excluded known neighbors. We confirm that this offset is present regardless of the UV profile used to obtain the photometry with EMphot.  Visual inspection of the residuals for several ($\sim 10$) $g < 14$ galaxies suggests that many have poor model fits attributable to fragmentation in the optical image segmentation process. However, at this time we cannot establish whether EMphot or pipeline photometry is ultimately more consistent with the SDSS photometry, since the latter is also affected by the fragmentation. We do not find these galaxies to have any systematic trends with respect to optical size (half-light radius), only with respect to $g$-band magnitude.

\subsection{Effects of blending}
\label{Section::EffectsOfBlending}


\begin{figure*}
    \centering
    \gridline{
          \fig{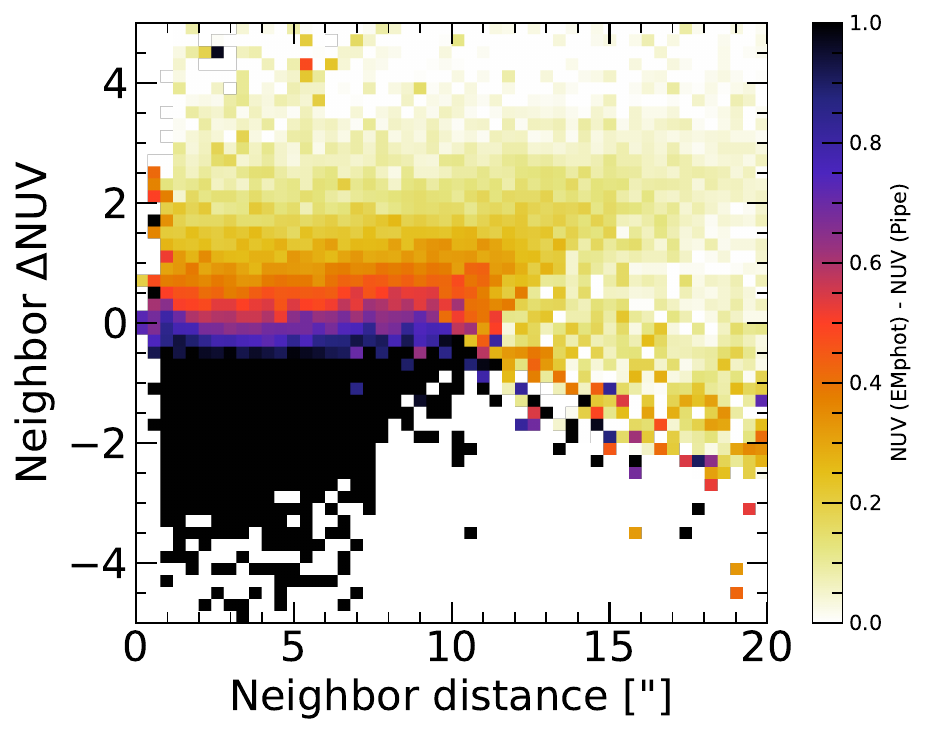}{0.45\textwidth}{}
          \fig{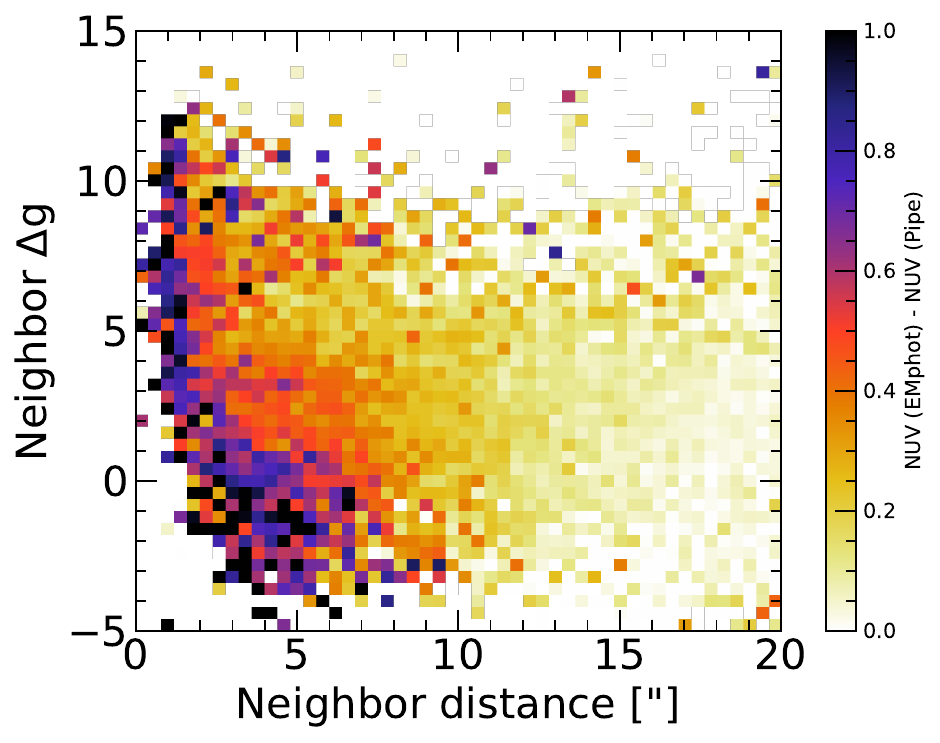}{0.45\textwidth}{}
          }
    \caption{2D histogram of the difference in EMphot NUV magnitude (left; neighbor NUV (EM) $-$ target NUV (EM), labeled $\Delta$NUV) and SDSS \textit{g} magnitude (right; neighbor \textit{g} $-$ target \textit{g}, labeled $\Delta$\textit{g}) between target and nearest neighbor versus their angular separation, colored by the difference in EMphot and pipeline magnitude of the target galaxy.  We limit the range of color to values between 0 and 1 to better highlight the various levels of blending.  Only galaxies with a single neighbor are shown.}
    \label{nnspace_deltacolor}
\end{figure*}

Next, we assess the fraction of galaxies with photometry significantly affected by blending.  We consider two regimes: moderate blends where the effect of blending in NUV is $> 0.2$ mag, and severe blends where the effect is $> 1$ mag. Determination of this fraction by selecting only by the difference in magnitude would be incorrect due to the random scatter in magnitudes.  We correct for the expected random scatter by subtracting the number of galaxies which scatter in the opposite direction.  For example, for moderate blends we compute the total blends as the number of galaxies with NUV (EM) $-$ NUV (Pipe) $> 0.2$ mag subtracted by the number of galaxies with NUV (EM) $-$ NUV (Pipe) $< -0.2$ mag. We find that among all target galaxies with NUV (Pipe) $\lesssim 22$, $16\%$ are moderately affected by blending ($> 0.2$ mag) and $2\%$ are severely affected ($> 1$ mag). 

Our next goal is to determine the regimes in which the effects of blending contamination of pipeline photometry are significant. This exercise may be of use when considering GALEX imaging not encompassed by this work (i.e., outside of SDSS DR10 footprint). In Figure \ref{nnspace_deltacolor} we show how the difference between NUV (EM) and NUV (Pipe) (i.e., the degree of contamination) varies as a function of two variables: (1) target vs. neighbor magnitude difference (left panel $\Delta$NUV (EMphot); right panel $\Delta$\textit{g}), and (2) distance between the target and the neighbor. Note that neighbors are still defined as object included in EMphot processing, i.e., those within 20\arcsec{} of a target and $u < 22$ mag.  Figure \ref{nnspace_deltacolor} shows only galaxies with a single neighbor within 20\arcsec{} to avoid confusion arising from the effects of multiple close neighbors.  We see that most of the significant blending (magnitude difference $\gtrsim 0.2$) occurs in the regime of separation $\lesssim$ 10\arcsec{}, roughly twice the GALEX FWHM, and when the neighbor is either brighter in NUV or up to 2 mag fainter. Severe contamination ($\gtrsim 1$ mag) results when the neighbor NUV is brighter. The degree of contamination is more strongly correlated with target/neighbor magnitude difference in NUV than with the distance. 

The degree of contamination is poorly correlated with the target/neighbor magnitude difference in $g$ band. Blending effects are severe out to 6\arcsec{} when the neighbor is brighter in $g$. Blending can be severe even if the neighbor is several mag fainter in \textit{g} than the target, but only at  distances $\gtrsim 2$\arcsec{}. 

The strong correlation between the degree of contamination and $\Delta$NUV implies that the closest neighbors are preferentially bluer in UV-optical color, suggesting that these neighbors are star-forming regions embedded in the target identified as separate objects by SDSS.  While the inclusion of fragments as separate objects may appear to be a weakness of our method compared to the GALEX pipeline, it actually ensures that the EMphot magnitudes are consistent with those of SDSS since SDSS model magnitudes will not include the contribution of these fragments either. 

\subsection{Consistency of photometry in the case of multiple detections in MIS and edge-of-detector bias}



\begin{figure}
    \centering
    \includegraphics[scale=0.53]{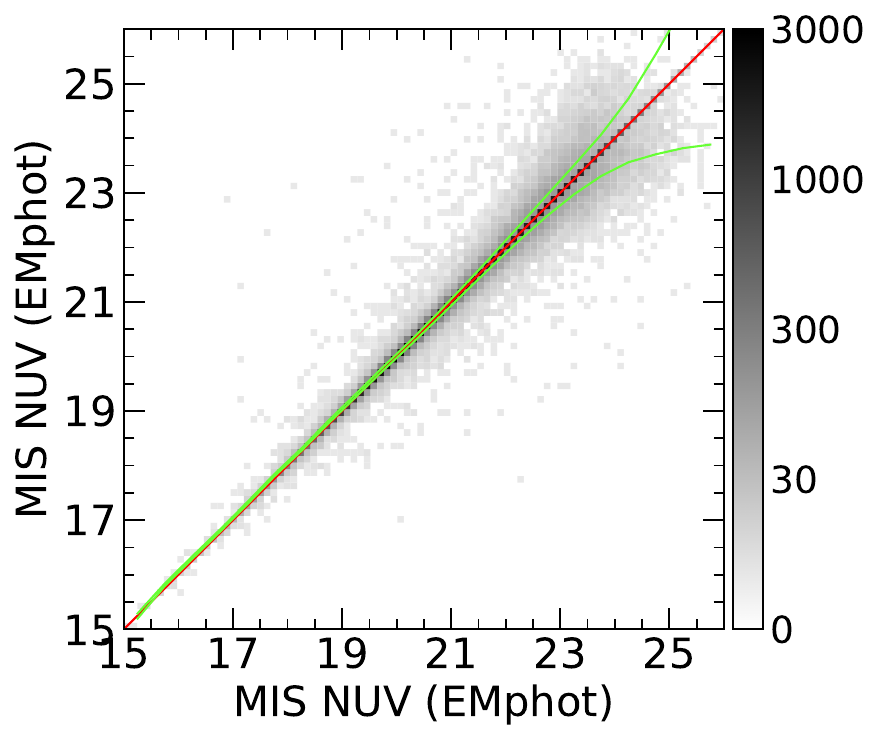}
    \caption{NUV magnitude comparison for target galaxies that are detected in multiple MIS tiles.  Magnitudes are on average consistent, i.e., there are no systematic offsets. The green lines represent the median magnitude $+/-$ the median error (at the given magnitude).}
    \label{multitile_NUVcomp}
\end{figure}

\begin{figure*}
    \centering
    \gridline{\fig{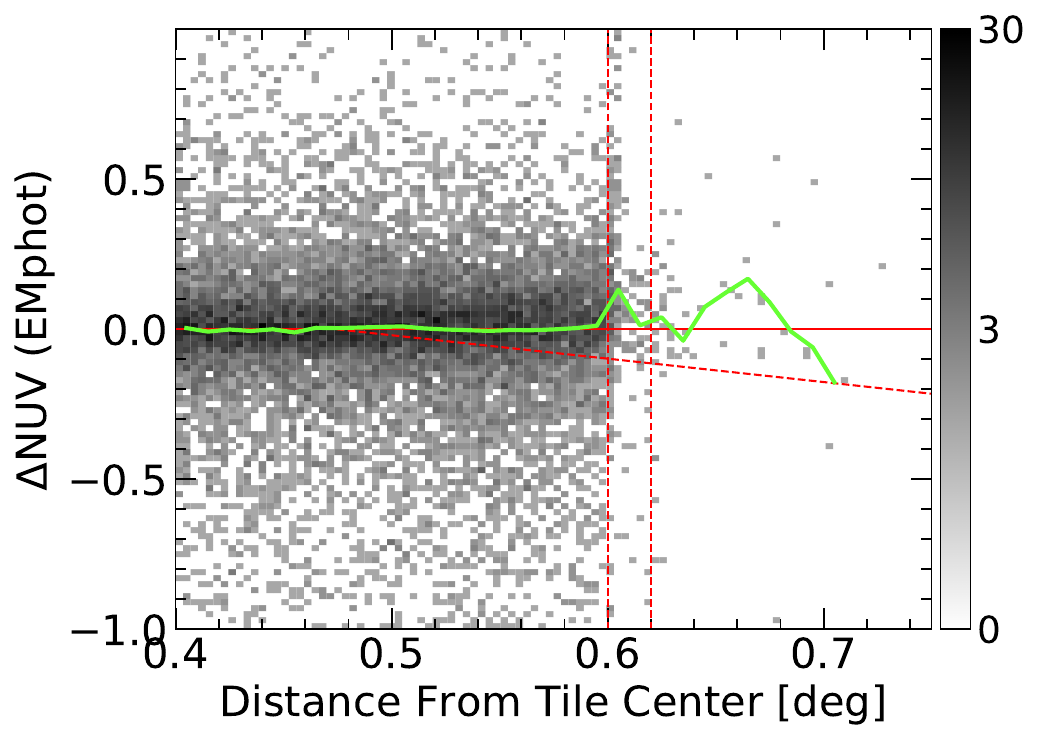}{0.45\textwidth}{}
          \fig{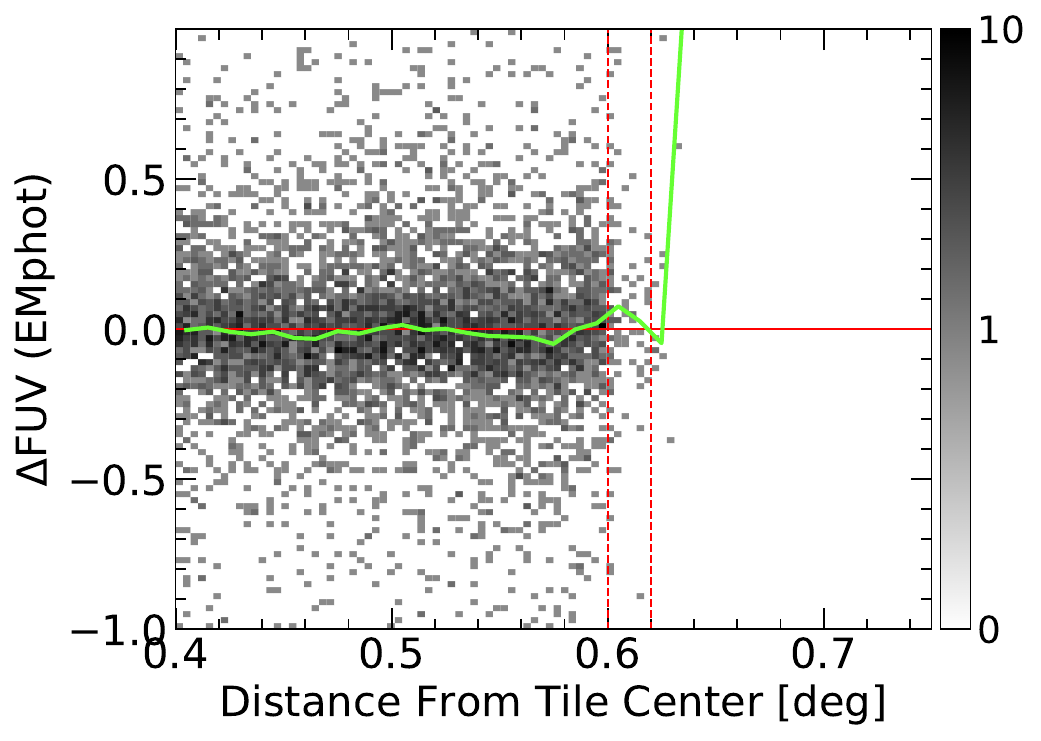}{0.45\textwidth}{}
          }
    \gridline{\fig{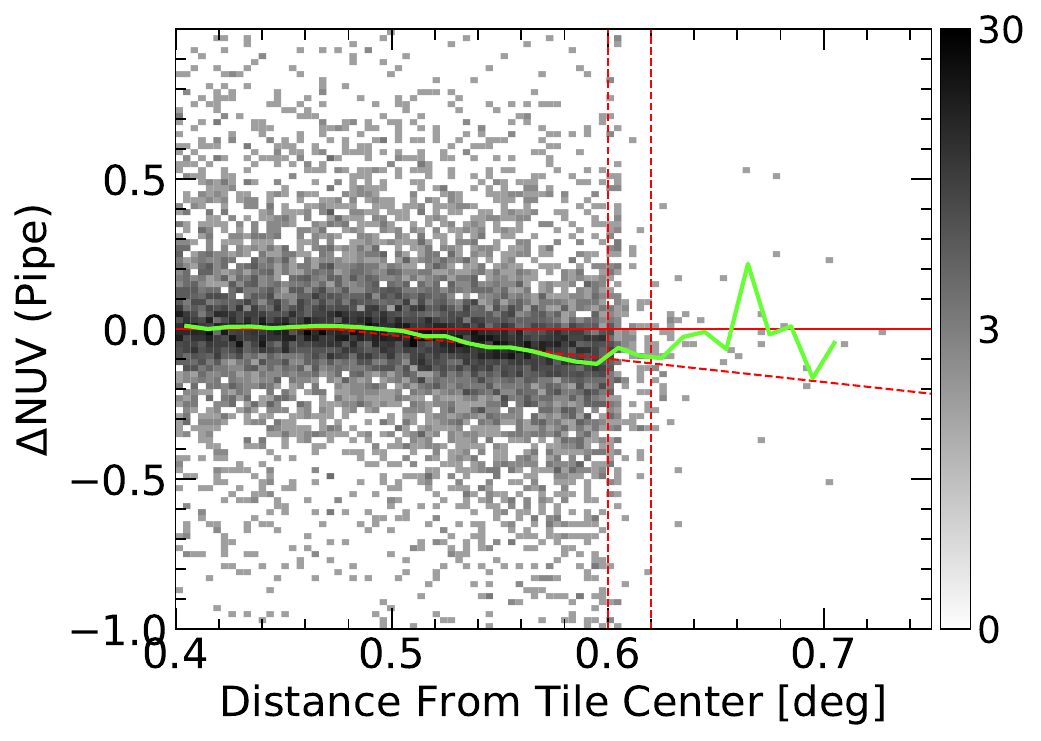}{0.45\textwidth}{}
          \fig{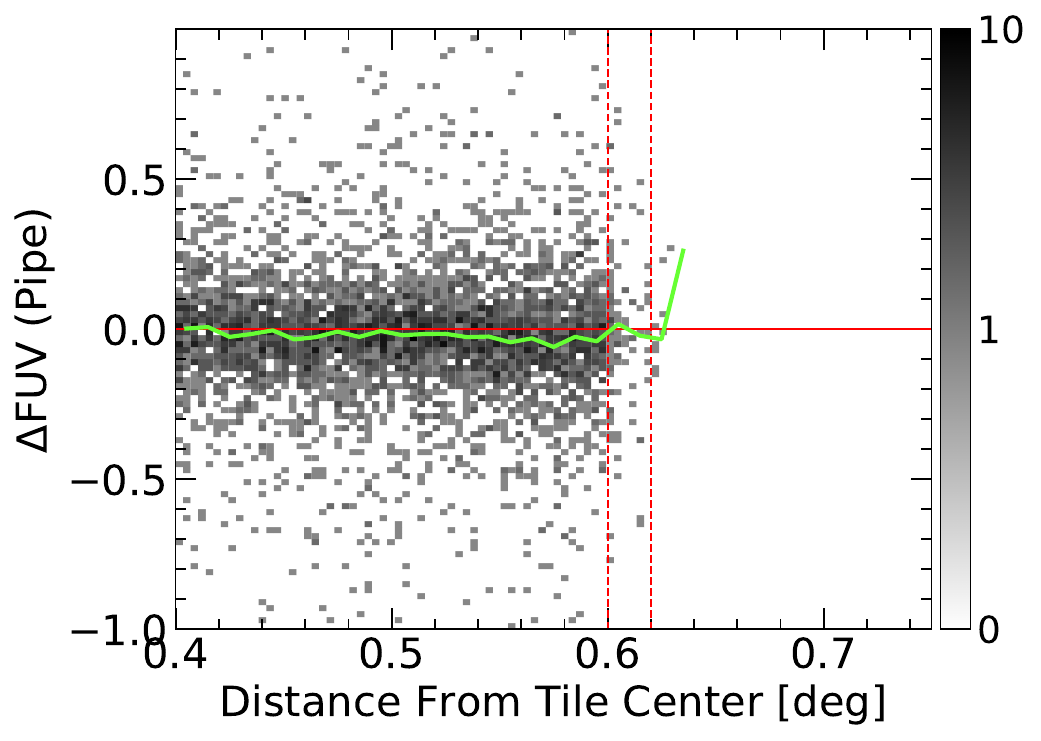}{0.45\textwidth}{}
          }
    \vspace*{-\baselineskip}
    \caption{Potential systematics in photometry as a function of the radial distance from image center. Figure shows the difference in EMphot (top row) and pipeline (bottom row) UV magnitude versus the radial distance from the tile center. Figure include targets that were imaged in two MIS tiles. The first magnitude in the difference is associated with the detection where the object appears at a larger distance (more subject to potential bias), and it is this distance that is shown on the x-axis.  We show the distance-dependent correction from \citet{Salim2016GSWLC} (red dashed line) as well as red dashed lines signifying detector edge (distance from center $=0.6$ and $0.62$ deg).  The pipeline NUV photometry suffers from a distance-dependent systematic, but not the photometry from EMphot. The green lines are running medians.}
    \label{multitile_FOV_plots}
\end{figure*}

Objects that lie close to a detector edge have biased NUV pipeline photometry, with magnitudes too bright by as much as 0.1 mag \citep{Salim2016GSWLC}.  Here we assess whether EMphot magnitudes are affected by the same distance (from tile center) dependent systematics. 

Among MIS targets, $23\%$ have multiple detections, with up to 10 detections for a given target galaxy.  For this analysis we only consider one duplicate observation per galaxy, which is chosen randomly from among the available detections.  We show the magnitude comparison for the pair of detections in Figure \ref{multitile_NUVcomp}.  As expected. there appear to be no systematic differences in magnitudes. Figure \ref{multitile_NUVcomp} also shows the median magnitude $+/-$ the median random error (at that magnitude) as green lines, which we find to be consistent with the observed scatter (i.e., consistent with $+/- 1$ standard deviation).

In Figure \ref{multitile_FOV_plots} we show the difference in NUV and FUV magnitudes versus the distance from tile center (in degrees), i.e., the distance from the tile center, for a subset of duplicate detections from Figure \ref{multitile_NUVcomp}.  To get $\Delta$NUV and $\Delta$FUV we subtract the magnitude of the detection with the smaller distance from the magnitude of a detection with a larger distance.  We show only detection pairs for which the larger distance is $> 0.4$ deg and the smaller distance is $< 0.4$ deg.  The idea is that in each such pair the detection with distance $<0.4$ deg will be unbiased and can serve as the ground truth. The $x$-axis represents the distance of the detection lying closer to the edge and thus subject to the potential bias.  The distance from tile center correction found by \citet{Salim2016GSWLC} (based on an entirely different method whereby the UV magnitude of a galaxy was crudely predicted from its optical colors) is shown in the left panels as a red dashed line. 

From Figure \ref{multitile_FOV_plots}, we can confirm that the pipeline photometry is biased at large distances from the tile center in the manner described by \citet{Salim2016GSWLC}.  In contrast, the EMphot photometry is free of distance-dependent systematics up to distance $= 0.6$ deg.  For a tiny region occupying $0.6 <$ distance $< 0.62$ deg (marked with dashed lines in the bottom panels) corresponding to the very edge of the detector, we find that the magnitudes are underestimated by $\sim 0.1$ mag on average. Detections that fall in this distance sliver are omitted from our final catalog.  Values of distance $> 0.62$ deg, which is well beyond the edge of the detector, are seemingly present. This is because in rare cases the co-addition of multiple visits involved tiles whose centers are off, and so represent detections with smaller distance in another tile center (e.g., distance $=0.7$ deg is actually distance $\leq0.5$ deg from another tile center).  Altogether EMphot offers an improvement over the pipeline in the edge of the detector regime by implicitly accounting for the systematics dependent on distance from the tile center.  

Furthermore, early GALEX processing contained spatial distortions close to the edge \citep{Morrissey2007GALEX}. These have been mitigated in more recent image processing and in any case have no effect on the photometry, as shown by the preceding analysis. 

\subsection{Assessment of calibration errors}

\begin{figure*}
    \centering
    \gridline{\fig{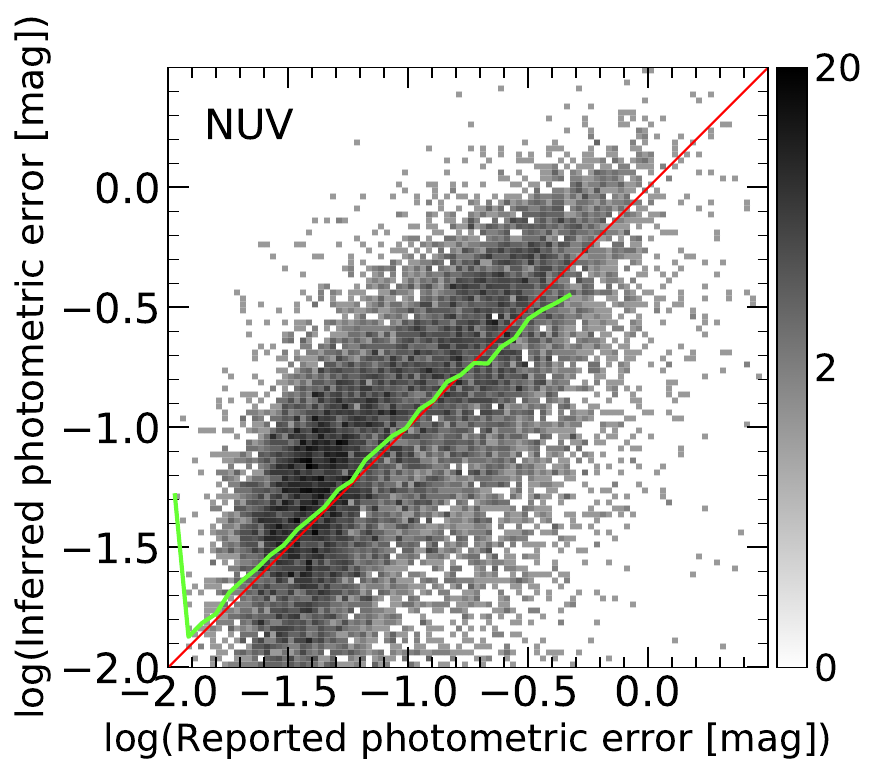}{0.45\textwidth}{}
          \fig{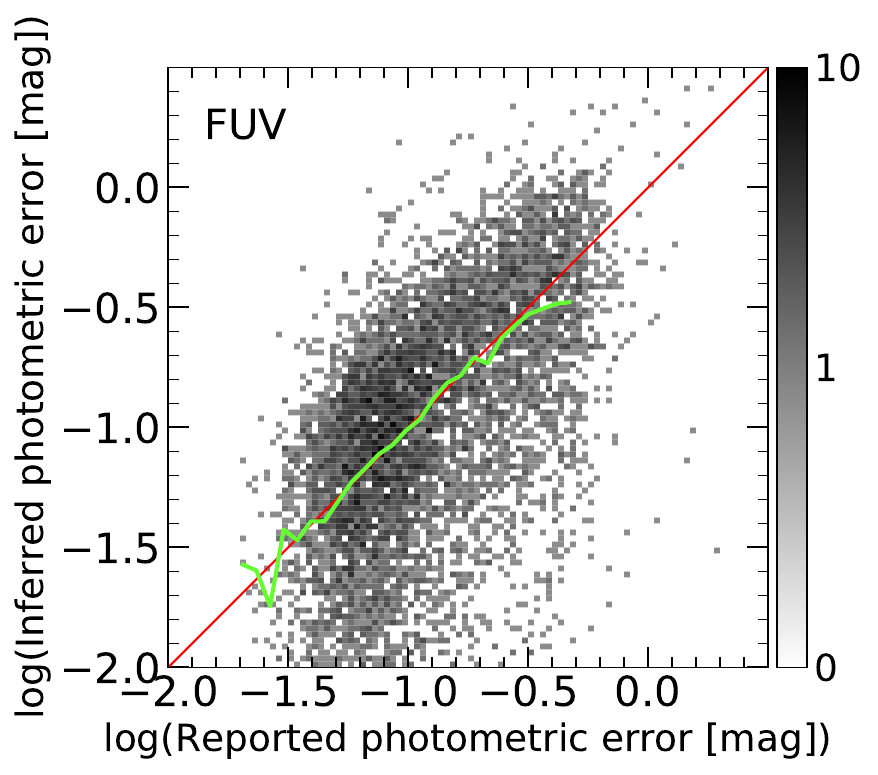}{0.45\textwidth}{}
          }
    \vspace*{-\baselineskip}
    \caption{Test of the robustness of EMphot photometry errors. Shown are inferred  versus EMphot-reported random photometric errors for galaxies with multiple detections in tiles of MIS depth. Only galaxies without neighbors are shown.  The inferred error is based on two drawings from a Gaussian, the $1\sigma$ width of which is given by $|m_1 - m_2| \times \sqrt{\pi} / 2$, while the reported random error is given by $\sqrt{ (\sigma_1^2 + \sigma_2^2) / 2 }$, with $m$ and $\sigma$ representing the EMphot magnitude and error, respectively. The agreement between the running median (green line) and 1:1 line (red line) means that the error reported by EMphot is robust.}
    \label{calibration_plots}
\end{figure*}

In this section we use duplicate observations to (1) assess the robustness of photometry errors reported by EMphot and to (2) constrain the degree of calibration errors. The idea behind the testing of the robustness of the reported error is the following.  If the error is given by the Gaussian with a standard deviation $\sigma$, then the difference of two drawings from that Gaussian will on average be $2 \sigma / \sqrt{\pi}$.  In other words, the true standard deviation (photometry error) can be inferred from the difference of two observations of the same galaxy and then compared to the photometric error reported by EMphot.  We note that for shallow imaging (e.g. AIS) the noise is more likely to be Poissonian than Gaussian, however for this assessment we use only MIS for which the noise is known to be Gaussian.  

Figure \ref{calibration_plots} shows the inferred photometry error versus the reported random error, in log scale, for NUV and FUV. Note that both errors are given in magnitude units, so each axis is in units of log (magnitude). We use the same sample as that used in Figure \ref{multitile_NUVcomp}, i.e., galaxies with multiple detections in tiles of MIS depth, but exclude galaxies with neighbors within 20\arcsec{} to avoid blending effects.  The inferred error is given by $|m_1 - m_2| \times \sqrt{\pi} / 2$, while the reported error is given by $\sqrt{ (\sigma_1^2 + \sigma_2^2) / 2 }$, where $m$ and $\sigma$ represent the EMphot magnitude and error, respectively.  The running median lines lie close to the 1:1 line, confirming the robustness of the errors reported by EMphot.

Note that the inferred error will contain both the random (photon shot counting) and any calibration error, whereas the EMphot reported error is only the random error. Calibration errors arise primarily from uncertainties in the flat field maps for different tiles and dominate over the nominal (photon shot counting) error only for bright sources with low photon shot noise. We note that the calibration error also includes any potential error in the GALEX zero points, though this error will be the same for all galaxies. The median trends of Figure \ref{calibration_plots} do not allow us to discern the calibration error given the paucity of bright objects, but we can infer an upper bound of inferred error of 0.015 mag in NUV and 0.025 mag in FUV, which is fully consistent with the flux calibration error determined by GALEX processing team of $0.8\%$ in NUV and $2.5\%$ in FUV\footnote{\url{http://www.galex.caltech.edu/researcher/gr6_docs/GI_Doc_Ops7.pdf}}. 

We have assumed that the magnitude error distributions for MIS targets are Gaussian.  To verify this assumption we perform a mock test. For each MIS target we draw twice from a Gaussian centered on the EMphot (NUV) magnitude with width equal to the magnitude error, and also draw twice from a Gaussian centered on the equivalent flux and flux error. We find that the error inferred from both the difference in magnitudes and the difference in fluxes is consistent to within $\sim3\%$. Thus we find it safe to assume Gaussian error distributions for MIS targets. 
Furthermore, any deviation from Gaussianity is so small that any impact on our estimate of the calibration error is negligible.

\subsection{Consistency of photometry between UV surveys}
\label{Section:UVsurveyComp}

\begin{figure*}
    \centering
    \gridline{\fig{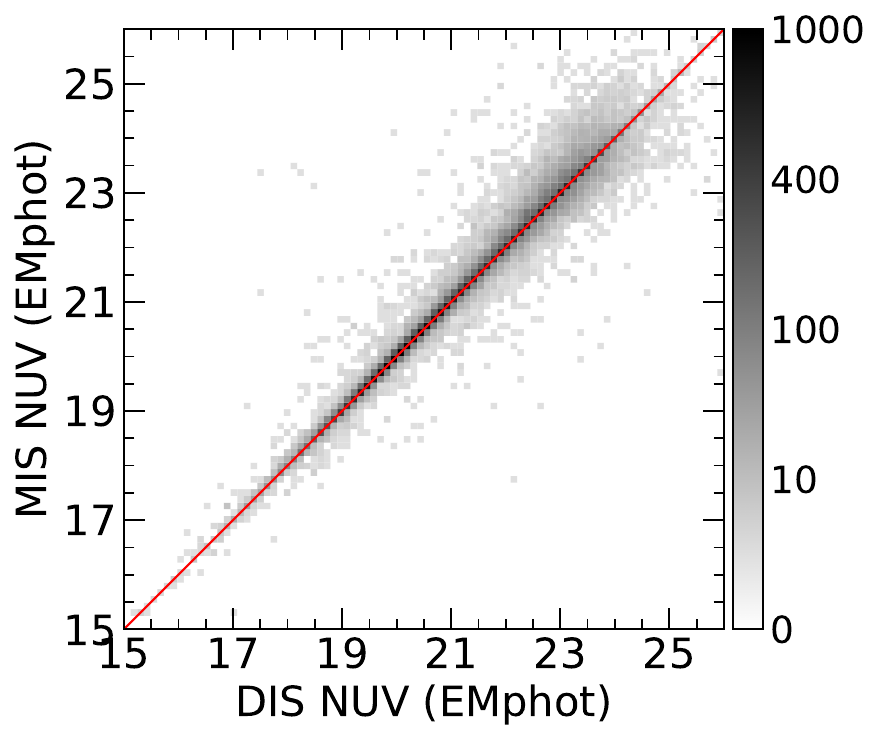}{0.3\textwidth}{}
          \fig{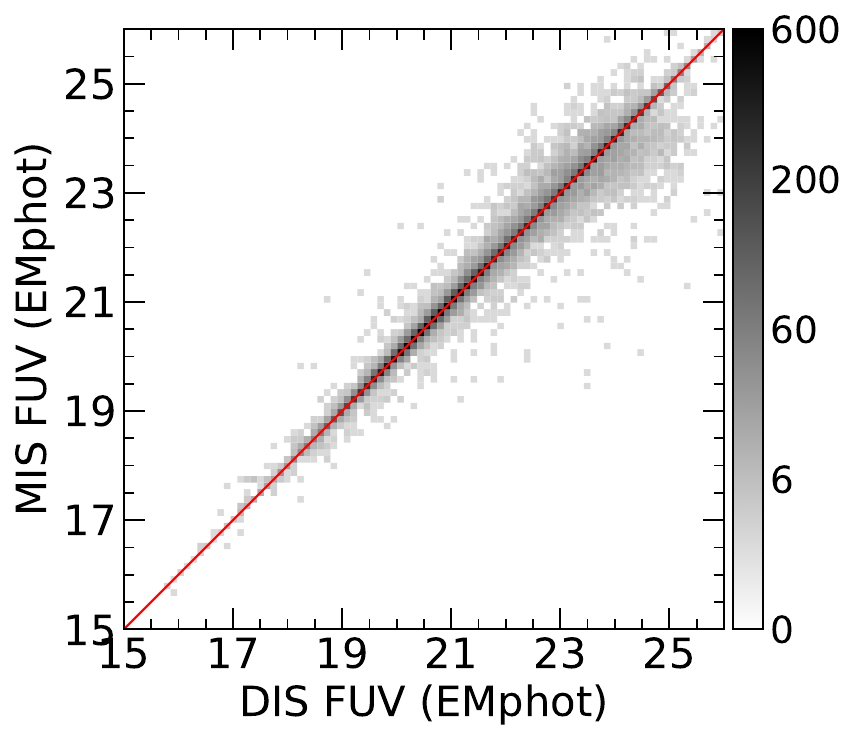}{0.3\textwidth}{}
          \fig{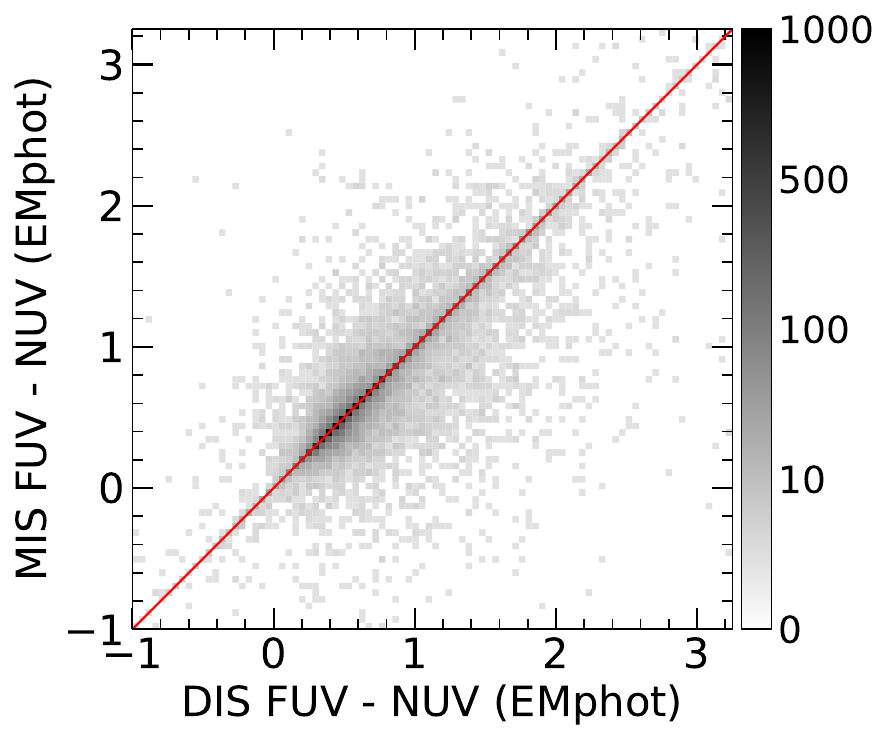}{0.3\textwidth}{}
          }
    \gridline{\fig{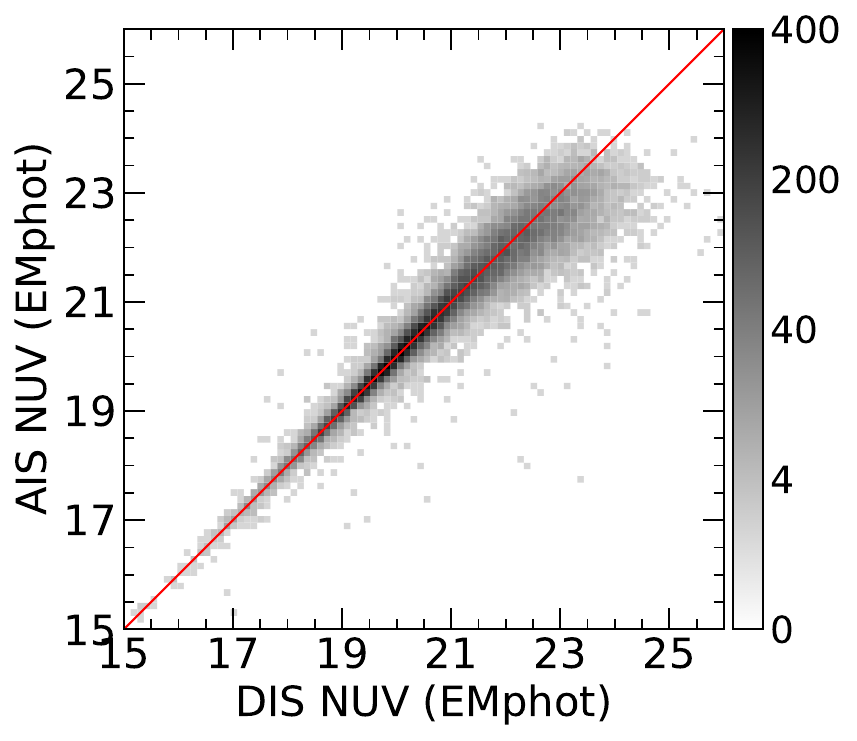}{0.3\textwidth}{}
          \fig{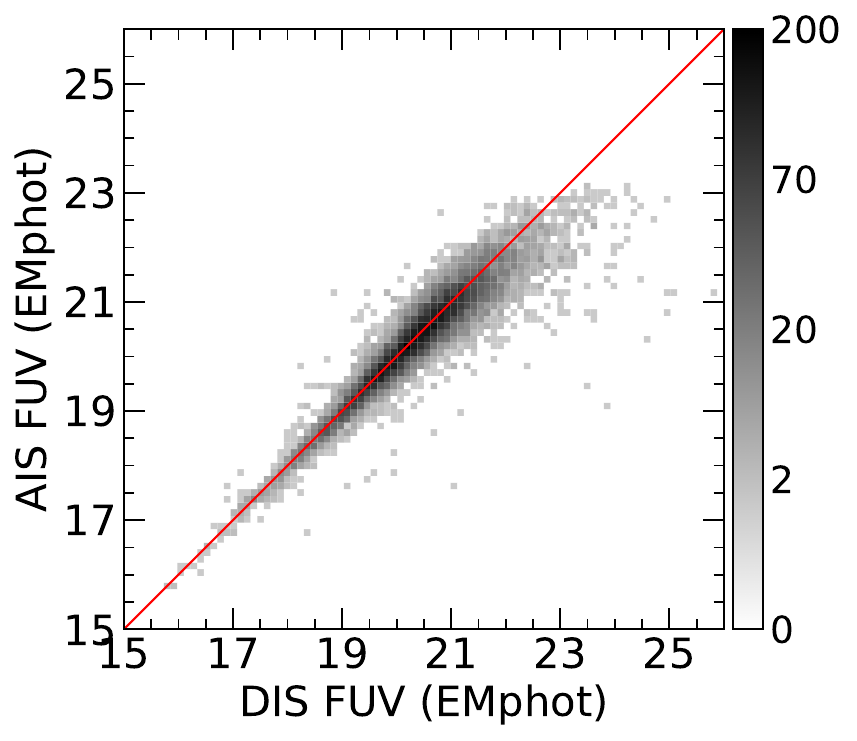}{0.3\textwidth}{}
          \fig{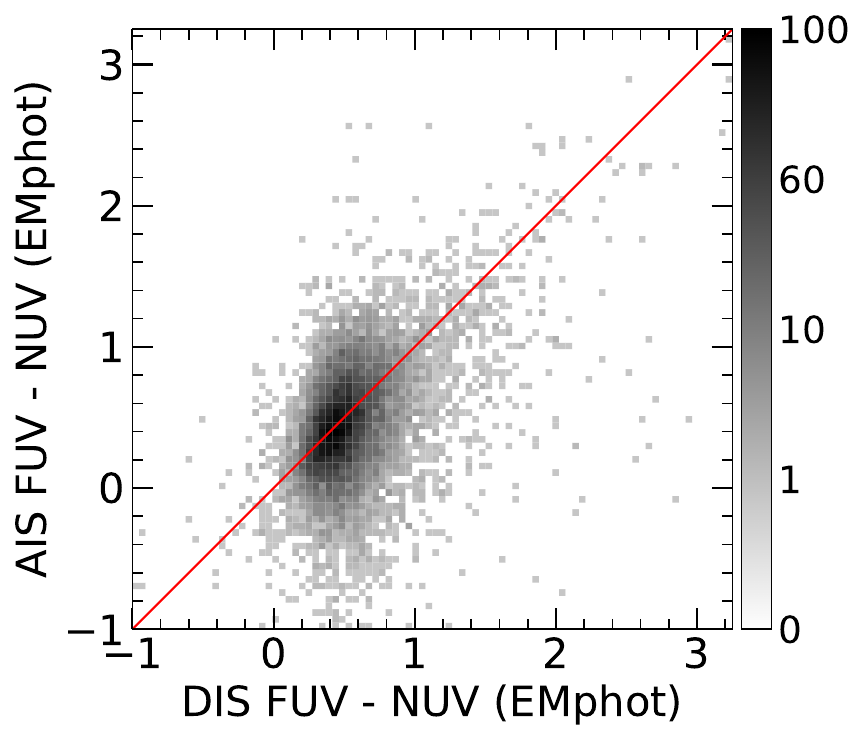}{0.3\textwidth}{}
          }
    \vspace*{-\baselineskip}
    \caption{Comparison of MIS (top row) and AIS (bottom row) NUV magnitudes (left column), FUV magnitudes (middle column), and UV colors (right column) versus DIS for target galaxies with detections in both surveys.  The shallower depth of AIS and MIS surveys with respect to DIS does not incur systematics, although for AIS there is a considerable scatter in UV colors.}
    \label{multisurvey_plots}
\end{figure*}

We explore the consistency of photometry between UV surveys of different depth (AIS, MIS and DIS) by considering DIS targets with detections in either AIS or MIS.  1:1 comparisons of EMphot magnitudes and colors between MIS and DIS are shown in the top row of Figure \ref{multisurvey_plots}.  We see that there are no systematic differences between MIS and DIS even for very faint objects.  We show a similar comparison between AIS and DIS in the bottom row of Figure \ref{multisurvey_plots}. There is a weak trend among the faintest galaxies where AIS magnitudes skew brighter than those of MIS or DIS; the trend is not observed between MIS and DIS, so it is attributable to limitations of the AIS photometry, possibly a difficulty in estimating the sky background which would more strongly impact the faintest sources. AIS UV colors show considerable scatter with respect to DIS colors but still no significant systematics.  Altogether we find that EMphot photometry is largely unbiased by depth.  

\subsection{Description of the EMphot photometry catalog}
\label{Section:FinalCatalog}

We compile the EMphot photometry for all target galaxies and release it in the form of several catalogs, one each for AIS, MIS, DIS, (i.e., corresponding to GSWLC-A, M and D lists) and one using the deepest available photometry (corresponding to GSWLC-X).  The catalog includes GALEX and SDSS object IDs in addition to the RA and DEC.  We further include the best EMphot magnitudes and errors for NUV and FUV which, as in GSWLC, in the case of multiple detections is associated with the tile with longest FUV exposure time (or longest NUV exposure time if no FUV image is available).  We also include the name of the tile and pipeline photometry.  If a galaxy has multiple observations within the same survey we give the number of such detections, the sum of exposure times, and the combined EMphot photometry (error-weighted averages).  Finally, we include a flag for the best model profile.  We show a list of columns and corresponding descriptions in Table \ref{SampleSizeTable}.  The combined photometry is preferable in general as it is the most precise.  However, the best-tile photometry may be preferred if one aims for measurements based on uniform depth, particularly for MIS.  We note that EMphot does not allow for negative fluxes by default.  Overall, $19\%$, $15\%$, and $33\%$ of DIS, MIS, and AIS targets (out of the total number in each GSWLC catalog), respectively, lack EMphot photometry. In some cases pipeline photometry for these objects may be present. In addition to the journal, the catalogs are publicly available at \url{https://salims.pages.iu.edu/galex}. 

\section{Conclusions}
\label{Section:Conclusion}

In this work we use the EMphot software coupled with synthetic images featuring various model profiles in order to perform forced GALEX photometry of $z < 0.3$ SDSS galaxies using optical prior positions of target galaxies and their neighbors.  The forced photometry offers an improvement over the original GALEX pipeline photometry, which is more susceptible to blending contamination due to the limited resolution of GALEX.  To account for the physical diversity in UV profiles, we separately fit three different model profiles: an optical-like profile, an exponential profile, and a flat profile.  For each galaxy's NUV photometry, we adopt the best-fitting profile.  The same best-fitting NUV model profile is then used to obtain the FUV photometry.  

Our use of optical priors and model photometry allows for not only robust deblending of sources but also offers other advantages.  To summarize:

\begin{enumerate}
\item EMphot provides NUV magnitudes for $79\%$ of galaxies originally without NUV magnitudes in the GALEX pipeline. 
\item The new magnitudes are on average consistent with the pipeline magnitudes for both FUV and NUV among sources with NUV $\lesssim 22$.  For fainter galaxies, EMphot predicts systematically brighter magnitudes (up to 0.5 mag), which we attribute to improved background estimation in EMphot.  
\item The new photometry reveals that $16\%$ of galaxies were moderately affected by blending in the pipeline photometry (contamination $>0.2$ mag) whereas $2.4\%$ were severely contaminated ($>1$ mag). 
\item No systematic differences in the model photometry between surveys of different depth (AIS, MIS, and DIS) are present. 
\item The new photometry does not suffer from edge-of-detector systematics which were biasing the NUV pipeline photometry by up to 0.1 mag.  
\item The catalog of EMphot forced GALEX photometry for all target galaxies is being made publicly available.  
\end{enumerate}

With more accurate UV photometry for $\sim700,000$ galaxies, one can expect to obtain more reliable physical parameters from the SED fitting, most crucially the star formation rates.  In the future, our methodological framework could also be applied to galaxies which lie outside the brightness and redshift cuts of the current samples, to better understand the UV properties of fainter or higher-redshift galaxies.  

The construction of GSWLC was funded through NASA awards NNX12AE06G and 80NSSC20K0440. The authors express their gratitude to the developers of the EMphot software. M\'ed\'eric Boquien acknowledges support from FONDECYT regular grant 1211000 and by the ANID BASAL project FB210003. 

\begin{deluxetable*}{rlll}[ht!]
\tablecaption{Contents of EMphot forced photometry catalogs.  Columns are the same for each catalog (GSWLC-A, M, D, X), corresponding to survey AIS, MIS, DIS, or the deepest of three (GSWLC-X). Electronic versions of the catalogs are available online at \url{https://salims.pages.iu.edu/galex}.}
\tablecolumns{22}
\tablenum{1}
\label{SampleSizeTable}
\tablewidth{0pt}
\tablehead{
\colhead{Column number} &
\colhead{Column name} & \colhead{Units} & \colhead{Description}
}
\startdata
1 & objID &  & SDSS photometric identification number  \\
2 & GLXID &  & GALEX photometric identification number  \\
3 & RA & deg & Right Ascension from SDSS  \\
4 & Decl. & deg & Declination from SDSS  \\
5 & Tile & & Name of GALEX tile with longest exposure time (best tile)\\
6 & nuv\textunderscore expt & seconds & NUV exposure time for best tile \\
7 & fuv\textunderscore expt & seconds & FUV exposure time for best tile \\
8 & bestRun & & Best model profile flag (0 = optical-like, 1 = exponential, 2 = flat) \\
9 & nuv\textunderscore em & mag & EMphot NUV magnitude for best tile \\
10 & nuv\textunderscore em\textunderscore err & mag & Error of EMphot NUV magnitude for best tile \\
11 & fuv\textunderscore em & mag & EMphot FUV magnitude for best tile \\
12 & fuv\textunderscore em\textunderscore err & mag & Error of EMphot FUV magnitude for best tile \\
13 & combined\textunderscore nuv\textunderscore expt & seconds & Sum of NUV exposure time for all detections \\
14 & combined\textunderscore fuv\textunderscore expt & seconds & Sum of FUV exposure time for all detections \\
15 & combined\textunderscore nuv\textunderscore em & mag & Combined EMphot NUV magnitude \\
16 & combined\textunderscore nuv\textunderscore em\textunderscore err & mag & Error of combined EMphot NUV magnitude \\
17 & combined\textunderscore fuv\textunderscore em & mag & Combined EMphot FUV magnitude \\
18 & combined\textunderscore fuv\textunderscore em\textunderscore err & mag &  Error of combined EMphot FUV magnitude \\
19 & ndetect\textunderscore nuv & & Number of NUV detections \\
20 & ndetect\textunderscore fuv & & Number of FUV detections \\
21 & nuv\textunderscore pipe & mag & GALEX pipeline NUV magnitude \\
22 & nuv\textunderscore pipe\textunderscore err & mag & Error of GALEX pipeline NUV magnitude \\
23 & fuv\textunderscore pipe & mag & GALEX pipeline FUV magnitude \\
24 & fuv\textunderscore pipe\textunderscore err & mag & Error of GALEX pipeline FUV magnitude \\
25 & nbg\textunderscore contam\textunderscore flag & & Flag indicating possible contamination of UV flux from a neighbor $>$20\arcsec{} \\
 & & & from the target; 0 = no contamination, 1 = possible contamination
\enddata
\end{deluxetable*}



\bibliography{Bibliography}

\begin{thebibliography}{}
\expandafter\ifx\csname natexlab\endcsname\relax\def\natexlab#1{#1}\fi
\providecommand{\url}[1]{\href{#1}{#1}}

\bibitem[{{Ahumada} {et~al.}(2020){Ahumada}, {Prieto}, {Almeida}, {Anders},
  {Anderson}, {Andrews}, {Anguiano}, {Arcodia}, {Armengaud}, {Aubert}, {Avila},
  {Avila-Reese}, {Badenes}, {Balland}, {Barger}, {Barrera-Ballesteros}, {Basu},
  {Bautista}, {Beaton}, {Beers}, {Benavides}, {Bender}, {Bernardi}, {Bershady},
  {Beutler}, {Bidin}, {Bird}, {Bizyaev}, {Blanc}, {Blanton}, {Boquien},
  {Borissova}, {Bovy}, {Brandt}, {Brinkmann}, {Brownstein}, {Bundy}, {Bureau},
  {Burgasser}, {Burtin}, {Cano-D{\'\i}az}, {Capasso}, {Cappellari}, {Carrera},
  {Chabanier}, {Chaplin}, {Chapman}, {Cherinka}, {Chiappini}, {Doohyun Choi},
  {Chojnowski}, {Chung}, {Clerc}, {Coffey}, {Comerford}, {Comparat}, {da
  Costa}, {Cousinou}, {Covey}, {Crane}, {Cunha}, {Ilha}, {Dai}, {Damsted},
  {Darling}, {Davidson}, {Davies}, {Dawson}, {De}, {de la Macorra}, {De Lee},
  {Queiroz}, {Deconto Machado}, {de la Torre}, {Dell'Agli}, {du Mas des
  Bourboux}, {Diamond-Stanic}, {Dillon}, {Donor}, {Drory}, {Duckworth},
  {Dwelly}, {Ebelke}, {Eftekharzadeh}, {Davis Eigenbrot}, {Elsworth},
  {Eracleous}, {Erfanianfar}, {Escoffier}, {Fan}, {Farr},
  {Fern{\'a}ndez-Trincado}, {Feuillet}, {Finoguenov}, {Fofie},
  {Fraser-McKelvie}, {Frinchaboy}, {Fromenteau}, {Fu}, {Galbany}, {Garcia},
  {Garc{\'\i}a-Hern{\'a}ndez}, {Oehmichen}, {Ge}, {Maia}, {Geisler}, {Gelfand},
  {Goddy}, {Gonzalez-Perez}, {Grabowski}, {Green}, {Grier}, {Guo}, {Guy},
  {Harding}, {Hasselquist}, {Hawken}, {Hayes}, {Hearty}, {Hekker}, {Hogg},
  {Holtzman}, {Horta}, {Hou}, {Hsieh}, {Huber}, {Hunt}, {Chitham}, {Imig},
  {Jaber}, {Angel}, {Johnson}, {Jones}, {J{\"o}nsson}, {Jullo}, {Kim},
  {Kinemuchi}, {Kirkpatrick}, {Kite}, {Klaene}, {Kneib}, {Kollmeier}, {Kong},
  {Kounkel}, {Krishnarao}, {Lacerna}, {Lan}, {Lane}, {Law}, {Le Goff}, {Leung},
  {Lewis}, {Li}, {Lian}, {Lin}, {Long}, {Longa-Pe{\~n}a}, {Lundgren}, {Lyke},
  {Ted Mackereth}, {MacLeod}, {Majewski}, {Manchado}, {Maraston}, {Martini},
  {Masseron}, {Masters}, {Mathur}, {McDermid}, {Merloni}, {Merrifield},
  {M{\'e}sz{\'a}ros}, {Miglio}, {Minniti}, {Minsley}, {Miyaji}, {Mohammad},
  {Mosser}, {Mueller}, {Muna}, {Mu{\~n}oz-Guti{\'e}rrez}, {Myers}, {Nadathur},
  {Nair}, {Nandra}, {do Nascimento}, {Nevin}, {Newman}, {Nidever}, {Nitschelm},
  {Noterdaeme}, {O'Connell}, {Olmstead}, {Oravetz}, {Oravetz}, {Osorio},
  {Pace}, {Padilla}, {Palanque-Delabrouille}, {Palicio}, {Pan}, {Pan},
  {Parker}, {Paviot}, {Peirani}, {Ram{\'r}ez}, {Penny}, {Percival},
  {Perez-Fournon}, {P{\'e}rez-R{\`a}fols}, {Petitjean}, {Pieri},
  {Pinsonneault}, {Poovelil}, {Povick}, {Prakash}, {Price-Whelan}, {Raddick},
  {Raichoor}, {Ray}, {Rembold}, {Rezaie}, {Riffel}, {Riffel}, {Rix}, {Robin},
  {Roman-Lopes}, {Rom{\'a}n-Z{\'u}{\~n}iga}, {Rose}, {Ross}, {Rossi},
  {Rowlands}, {Rubin}, {Salvato}, {S{\'a}nchez}, {S{\'a}nchez-Menguiano},
  {S{\'a}nchez-Gallego}, {Sayres}, {Schaefer}, {Schiavon}, {Schimoia},
  {Schlafly}, {Schlegel}, {Schneider}, {Schultheis}, {Schwope}, {Seo},
  {Serenelli}, {Shafieloo}, {Shamsi}, {Shao}, {Shen}, {Shetrone}, {Shirley},
  {Aguirre}, {Simon}, {Skrutskie}, {Slosar}, {Smethurst}, {Sobeck}, {Sodi},
  {Souto}, {Stark}, {Stassun}, {Steinmetz}, {Stello}, {Stermer},
  {Storchi-Bergmann}, {Streblyanska}, {Stringfellow}, {Stutz}, {Su{\'a}rez},
  {Sun}, {Taghizadeh-Popp}, {Talbot}, {Tayar}, {Thakar}, {Theriault}, {Thomas},
  {Thomas}, {Tinker}, {Tojeiro}, {Toledo}, {Tremonti}, {Troup}, {Tuttle},
  {Unda-Sanzana}, {Valentini}, {Vargas-Gonz{\'a}lez}, {Vargas-Maga{\~n}a},
  {V{\'a}zquez-Mata}, {Vivek}, {Wake}, {Wang}, {Weaver}, {Weijmans}, {Wild},
  {Wilson}, {Wilson}, {Wolthuis}, {Wood-Vasey}, {Yan}, {Yang}, {Y{\`e}che},
  {Zamora}, {Zarrouk}, {Zasowski}, {Zhang}, {Zhao}, {Zhao}, {Zheng}, {Zheng},
  {Zhu}, \& {Zou}}]{Ahumada2020SDSSDR16datarelease}
{Ahumada}, R., {Prieto}, C.~A., {Almeida}, A., {et~al.} 2020, \apjs, 249, 3

\bibitem[{{Astropy Collaboration} {et~al.}(2013){Astropy Collaboration},
  {Robitaille}, {Tollerud}, {Greenfield}, {Droettboom}, {Bray}, {Aldcroft},
  {Davis}, {Ginsburg}, {Price-Whelan}, {Kerzendorf}, {Conley}, {Crighton},
  {Barbary}, {Muna}, {Ferguson}, {Grollier}, {Parikh}, {Nair}, {Unther},
  {Deil}, {Woillez}, {Conseil}, {Kramer}, {Turner}, {Singer}, {Fox}, {Weaver},
  {Zabalza}, {Edwards}, {Azalee Bostroem}, {Burke}, {Casey}, {Crawford},
  {Dencheva}, {Ely}, {Jenness}, {Labrie}, {Lim}, {Pierfederici}, {Pontzen},
  {Ptak}, {Refsdal}, {Servillat}, \& {Streicher}}]{Astropy2013}
{Astropy Collaboration}, {Robitaille}, T.~P., {Tollerud}, E.~J., {et~al.} 2013,
  \aap, 558, A33

\bibitem[{{Astropy Collaboration} {et~al.}(2018){Astropy Collaboration},
  {Price-Whelan}, {Sip{\H{o}}cz}, {G{\"u}nther}, {Lim}, {Crawford}, {Conseil},
  {Shupe}, {Craig}, {Dencheva}, {Ginsburg}, {Vand erPlas}, {Bradley},
  {P{\'e}rez-Su{\'a}rez}, {de Val-Borro}, {Aldcroft}, {Cruz}, {Robitaille},
  {Tollerud}, {Ardelean}, {Babej}, {Bach}, {Bachetti}, {Bakanov}, {Bamford},
  {Barentsen}, {Barmby}, {Baumbach}, {Berry}, {Biscani}, {Boquien}, {Bostroem},
  {Bouma}, {Brammer}, {Bray}, {Breytenbach}, {Buddelmeijer}, {Burke},
  {Calderone}, {Cano Rodr{\'\i}guez}, {Cara}, {Cardoso}, {Cheedella}, {Copin},
  {Corrales}, {Crichton}, {D'Avella}, {Deil}, {Depagne}, {Dietrich}, {Donath},
  {Droettboom}, {Earl}, {Erben}, {Fabbro}, {Ferreira}, {Finethy}, {Fox},
  {Garrison}, {Gibbons}, {Goldstein}, {Gommers}, {Greco}, {Greenfield},
  {Groener}, {Grollier}, {Hagen}, {Hirst}, {Homeier}, {Horton}, {Hosseinzadeh},
  {Hu}, {Hunkeler}, {Ivezi{\'c}}, {Jain}, {Jenness}, {Kanarek}, {Kendrew},
  {Kern}, {Kerzendorf}, {Khvalko}, {King}, {Kirkby}, {Kulkarni}, {Kumar},
  {Lee}, {Lenz}, {Littlefair}, {Ma}, {Macleod}, {Mastropietro}, {McCully},
  {Montagnac}, {Morris}, {Mueller}, {Mumford}, {Muna}, {Murphy}, {Nelson},
  {Nguyen}, {Ninan}, {N{\"o}the}, {Ogaz}, {Oh}, {Parejko}, {Parley}, {Pascual},
  {Patil}, {Patil}, {Plunkett}, {Prochaska}, {Rastogi}, {Reddy Janga},
  {Sabater}, {Sakurikar}, {Seifert}, {Sherbert}, {Sherwood-Taylor}, {Shih},
  {Sick}, {Silbiger}, {Singanamalla}, {Singer}, {Sladen}, {Sooley},
  {Sornarajah}, {Streicher}, {Teuben}, {Thomas}, {Tremblay}, {Turner},
  {Terr{\'o}n}, {van Kerkwijk}, {de la Vega}, {Watkins}, {Weaver}, {Whitmore},
  {Woillez}, {Zabalza}, \& {Astropy Contributors}}]{Astropy2018}
{Astropy Collaboration}, {Price-Whelan}, A.~M., {Sip{\H{o}}cz}, B.~M., {et~al.}
  2018, \aj, 156, 123

\bibitem[{{Bertin} \& {Arnouts}(1996)}]{Bertin&Arnouts1996SExtractor}
{Bertin}, E., \& {Arnouts}, S. 1996, \aaps, 117, 393

\bibitem[{{Binggeli} \&
  {Cameron}(1991)}]{Binggeli1991SurfaceBrightnessProfiles}
{Binggeli}, B., \& {Cameron}, L.~M. 1991, \aap, 252, 27

\bibitem[{{Ciambur}(2016)}]{Ciambur2016Profiler}
{Ciambur}, B.~C. 2016, \pasa, 33, e062

\bibitem[{{Conroy}(2013)}]{Conroy2013SEDFittingReview}
{Conroy}, C. 2013, \araa, 51, 393

\bibitem[{{Conseil} {et~al.}(2011){Conseil}, {Vibert}, {Amouts}, {Milliard},
  {Zamojski}, {Liebaria}, \& {Guillaume}}]{Conseil2011EMphot}
{Conseil}, S., {Vibert}, D., {Amouts}, S., {et~al.} 2011, in Astronomical
  Society of the Pacific Conference Series, Vol. 442, Astronomical Data
  Analysis Software and Systems XX, ed. I.~N. {Evans}, A.~{Accomazzi}, D.~J.
  {Mink}, \& A.~H. {Rots}, 107

\bibitem[{{Dale} {et~al.}(2009){Dale}, {Cohen}, {Johnson}, {Schuster},
  {Calzetti}, {Engelbracht}, {Gil de Paz}, {Kennicutt}, {Lee}, {Begum},
  {Block}, {Dalcanton}, {Funes}, {Gordon}, {Johnson}, {Marble}, {Sakai},
  {Skillman}, {van Zee}, {Walter}, {Weisz}, {Williams}, {Wu}, \&
  {Wu}}]{Dale2009SpitzerLVLSurvey}
{Dale}, D.~A., {Cohen}, S.~A., {Johnson}, L.~C., {et~al.} 2009, \apj, 703, 517

\bibitem[{{Dey} {et~al.}(2019){Dey}, {Schlegel}, {Lang}, {Blum}, {Burleigh},
  {Fan}, {Findlay}, {Finkbeiner}, {Herrera}, {Juneau}, {Landriau}, {Levi},
  {McGreer}, {Meisner}, {Myers}, {Moustakas}, {Nugent}, {Patej}, {Schlafly},
  {Walker}, {Valdes}, {Weaver}, {Y{\`e}che}, {Zou}, {Zhou}, {Abareshi},
  {Abbott}, {Abolfathi}, {Aguilera}, {Alam}, {Allen}, {Alvarez}, {Annis},
  {Ansarinejad}, {Aubert}, {Beechert}, {Bell}, {BenZvi}, {Beutler}, {Bielby},
  {Bolton}, {Brice{\~n}o}, {Buckley-Geer}, {Butler}, {Calamida}, {Carlberg},
  {Carter}, {Casas}, {Castander}, {Choi}, {Comparat}, {Cukanovaite}, {Delubac},
  {DeVries}, {Dey}, {Dhungana}, {Dickinson}, {Ding}, {Donaldson}, {Duan},
  {Duckworth}, {Eftekharzadeh}, {Eisenstein}, {Etourneau}, {Fagrelius},
  {Farihi}, {Fitzpatrick}, {Font-Ribera}, {Fulmer}, {G{\"a}nsicke},
  {Gaztanaga}, {George}, {Gerdes}, {Gontcho}, {Gorgoni}, {Green}, {Guy},
  {Harmer}, {Hernandez}, {Honscheid}, {Huang}, {James}, {Jannuzi}, {Jiang},
  {Joyce}, {Karcher}, {Karkar}, {Kehoe}, {Kneib}, {Kueter-Young}, {Lan},
  {Lauer}, {Le Guillou}, {Le Van Suu}, {Lee}, {Lesser}, {Perreault Levasseur},
  {Li}, {Mann}, {Marshall}, {Mart{\'\i}nez-V{\'a}zquez}, {Martini}, {du Mas des
  Bourboux}, {McManus}, {Meier}, {M{\'e}nard}, {Metcalfe},
  {Mu{\~n}oz-Guti{\'e}rrez}, {Najita}, {Napier}, {Narayan}, {Newman}, {Nie},
  {Nord}, {Norman}, {Olsen}, {Paat}, {Palanque-Delabrouille}, {Peng},
  {Poppett}, {Poremba}, {Prakash}, {Rabinowitz}, {Raichoor}, {Rezaie},
  {Robertson}, {Roe}, {Ross}, {Ross}, {Rudnick}, {Safonova}, {Saha},
  {S{\'a}nchez}, {Savary}, {Schweiker}, {Scott}, {Seo}, {Shan}, {Silva},
  {Slepian}, {Soto}, {Sprayberry}, {Staten}, {Stillman}, {Stupak}, {Summers},
  {Sien Tie}, {Tirado}, {Vargas-Maga{\~n}a}, {Vivas}, {Wechsler}, {Williams},
  {Yang}, {Yang}, {Yapici}, {Zaritsky}, {Zenteno}, {Zhang}, {Zhang}, {Zhou}, \&
  {Zhou}}]{Dey2019LegacySurvey}
{Dey}, A., {Schlegel}, D.~J., {Lang}, D., {et~al.} 2019, \aj, 157, 168

\bibitem[{{Gil de Paz} {et~al.}(2007){Gil de Paz}, {Boissier}, {Madore},
  {Seibert}, {Joe}, {Boselli}, {Wyder}, {Thilker}, {Bianchi}, {Rey}, {Rich},
  {Barlow}, {Conrow}, {Forster}, {Friedman}, {Martin}, {Morrissey}, {Neff},
  {Schiminovich}, {Small}, {Donas}, {Heckman}, {Lee}, {Milliard}, {Szalay}, \&
  {Yi}}]{GildePaz2007GALEXUVAtlas}
{Gil de Paz}, A., {Boissier}, S., {Madore}, B.~F., {et~al.} 2007, \apjs, 173,
  185

\bibitem[{{Guillaume} {et~al.}(2006){Guillaume}, {Llebaria}, {Aymeric},
  {Arnouts}, \& {Milliard}}]{Guillaume2006GALEXEMphotDeblendingDeepFields}
{Guillaume}, M., {Llebaria}, A., {Aymeric}, D., {Arnouts}, S., \& {Milliard},
  B. 2006, in Society of Photo-Optical Instrumentation Engineers (SPIE)
  Conference Series, Vol. 6064, Image Processing: Algorithms and Systems,
  Neural Networks, and Machine Learning, ed. E.~R. {Dougherty}, J.~T. {Astola},
  K.~O. {Egiazarian}, N.~M. {Nasrabadi}, \& S.~A. {Rizvi}, 332--341

\bibitem[{{Ilbert} {et~al.}(2009){Ilbert}, {Capak}, {Salvato}, {Aussel},
  {McCracken}, {Sanders}, {Scoville}, {Kartaltepe}, {Arnouts}, {Le Floc'h},
  {Mobasher}, {Taniguchi}, {Lamareille}, {Leauthaud}, {Sasaki}, {Thompson},
  {Zamojski}, {Zamorani}, {Bardelli}, {Bolzonella}, {Bongiorno}, {Brusa},
  {Caputi}, {Carollo}, {Contini}, {Cook}, {Coppa}, {Cucciati}, {de la Torre},
  {de Ravel}, {Franzetti}, {Garilli}, {Hasinger}, {Iovino}, {Kampczyk},
  {Kneib}, {Knobel}, {Kovac}, {Le Borgne}, {Le Brun}, {Le F{\`e}vre}, {Lilly},
  {Looper}, {Maier}, {Mainieri}, {Mellier}, {Mignoli}, {Murayama}, {Pell{\`o}},
  {Peng}, {P{\'e}rez-Montero}, {Renzini}, {Ricciardelli}, {Schiminovich},
  {Scodeggio}, {Shioya}, {Silverman}, {Surace}, {Tanaka}, {Tasca}, {Tresse},
  {Vergani}, \& {Zucca}}]{Ilbert2009COSMOSphotoZ}
{Ilbert}, O., {Capak}, P., {Salvato}, M., {et~al.} 2009, \apj, 690, 1236

\bibitem[{{Lang} {et~al.}(2016{\natexlab{a}}){Lang}, {Hogg}, \&
  {Mykytyn}}]{Lang2016TheTractor}
{Lang}, D., {Hogg}, D.~W., \& {Mykytyn}, D. 2016{\natexlab{a}}, {The Tractor:
  Probabilistic astronomical source detection and measurement}, Astrophysics
  Source Code Library, record ascl:1604.008, , , ascl:1604.008

\bibitem[{{Lang} {et~al.}(2016{\natexlab{b}}){Lang}, {Hogg}, \&
  {Schlegel}}]{Lang2016unWISEforcedphotometry}
{Lang}, D., {Hogg}, D.~W., \& {Schlegel}, D.~J. 2016{\natexlab{b}}, \aj, 151,
  36

\bibitem[{{Llebaria} {et~al.}(2008){Llebaria}, {Magnelli}, {Arnouts}, {Pollo},
  {Milliard}, \& {Guillaume}}]{Llebaria2008SPIEGALEXEMphotOpticalPriors}
{Llebaria}, A., {Magnelli}, B., {Arnouts}, S., {et~al.} 2008, in Society of
  Photo-Optical Instrumentation Engineers (SPIE) Conference Series, Vol. 6812,
  Image Processing: Algorithms and Systems VI, ed. J.~T. {Astola}, K.~O.
  {Egiazarian}, \& E.~R. {Dougherty}, 68121F

\bibitem[{{Marcum} {et~al.}(2001){Marcum}, {O'Connell}, {Fanelli}, {Cornett},
  {Waller}, {Bohlin}, {Neff}, {Roberts}, {Smith}, {Cheng}, {Collins},
  {Hennessy}, {Hill}, {Hill}, {Hintzen}, {Landsman}, {Ohl}, {Parise}, {Smith},
  {Freedman}, {Kuchinski}, {Madore}, {Angione}, {Palma}, {Talbert}, \&
  {Stecher}}]{Marcum2001UVAtlas&Morphology}
{Marcum}, P.~M., {O'Connell}, R.~W., {Fanelli}, M.~N., {et~al.} 2001, \apjs,
  132, 129

\bibitem[{{Merlin} {et~al.}(2016){Merlin}, {Bourne}, {Castellano}, {Ferguson},
  {Wang}, {Derriere}, {Dunlop}, {Elbaz}, \& {Fontana}}]{Merlin2016TPHOT}
{Merlin}, E., {Bourne}, N., {Castellano}, M., {et~al.} 2016, \aap, 595, A97

\bibitem[{{Morrissey} {et~al.}(2007){Morrissey}, {Conrow}, {Barlow}, {Small},
  {Seibert}, {Wyder}, {Budav{\'a}ri}, {Arnouts}, {Friedman}, {Forster},
  {Martin}, {Neff}, {Schiminovich}, {Bianchi}, {Donas}, {Heckman}, {Lee},
  {Madore}, {Milliard}, {Rich}, {Szalay}, {Welsh}, \&
  {Yi}}]{Morrissey2007GALEX}
{Morrissey}, P., {Conrow}, T., {Barlow}, T.~A., {et~al.} 2007, \apjs, 173, 682

\bibitem[{{Moustakas} {et~al.}(2013){Moustakas}, {Coil}, {Aird}, {Blanton},
  {Cool}, {Eisenstein}, {Mendez}, {Wong}, {Zhu}, \&
  {Arnouts}}]{Moustakas2013PRIMUSConstraintsOnQuenchingMergingAndSMF}
{Moustakas}, J., {Coil}, A.~L., {Aird}, J., {et~al.} 2013, \apj, 767, 50

\bibitem[{{Moutard} {et~al.}(2016){Moutard}, {Arnouts}, {Ilbert}, {Coupon},
  {Hudelot}, {Vibert}, {Comte}, {Conseil}, {Davidzon}, {Guzzo}, {Llebaria},
  {Martin}, {McCracken}, {Milliard}, {Morrison}, {Schiminovich}, {Treyer}, \&
  {Van Werbaeke}}]{EMphotMoutardCFHTLS2016}
{Moutard}, T., {Arnouts}, S., {Ilbert}, O., {et~al.} 2016, \aap, 590, A102

\bibitem[{{Salim} {et~al.}(2018){Salim}, {Boquien}, \&
  {Lee}}]{Salim2018DustAttCurves}
{Salim}, S., {Boquien}, M., \& {Lee}, J.~C. 2018, \apj, 859, 11

\bibitem[{{Salim} {et~al.}(2009){Salim}, {Dickinson}, {Michael Rich},
  {Charlot}, {Lee}, {Schiminovich}, {P{\'e}rez-Gonz{\'a}lez}, {Ashby},
  {Papovich}, {Faber}, {Ivison}, {Frayer}, {Walton}, {Weiner}, {Chary},
  {Bundy}, {Noeske}, \& {Koekemoer}}]{Salim2009MidIRDeepFields}
{Salim}, S., {Dickinson}, M., {Michael Rich}, R., {et~al.} 2009, \apj, 700, 161

\bibitem[{{Salim} {et~al.}(2016){Salim}, {Lee}, {Janowiecki}, {da Cunha},
  {Dickinson}, {Boquien}, {Burgarella}, {Salzer}, \&
  {Charlot}}]{Salim2016GSWLC}
{Salim}, S., {Lee}, J.~C., {Janowiecki}, S., {et~al.} 2016, \apjs, 227, 2

\bibitem[{{S{\'e}rsic}(1963)}]{Sersic1963Profile}
{S{\'e}rsic}, J.~L. 1963, Boletin de la Asociacion Argentina de Astronomia La
  Plata Argentina, 6, 41

\bibitem[{{Sheth} {et~al.}(2010){Sheth}, {Regan}, {Hinz}, {Gil de Paz},
  {Men{\'e}ndez-Delmestre}, {Mu{\~n}oz-Mateos}, {Seibert}, {Kim},
  {Laurikainen}, {Salo}, {Gadotti}, {Laine}, {Mizusawa}, {Armus},
  {Athanassoula}, {Bosma}, {Buta}, {Capak}, {Jarrett}, {Elmegreen},
  {Elmegreen}, {Knapen}, {Koda}, {Helou}, {Ho}, {Madore}, {Masters},
  {Mobasher}, {Ogle}, {Peng}, {Schinnerer}, {Surace}, {Zaritsky},
  {Comer{\'o}n}, {de Swardt}, {Meidt}, {Kasliwal}, \&
  {Aravena}}]{Sheth2010S4GInitialRelease}
{Sheth}, K., {Regan}, M., {Hinz}, J.~L., {et~al.} 2010, \pasp, 122, 1397

\bibitem[{{Simard} {et~al.}(2011){Simard}, {Mendel}, {Patton}, {Ellison}, \&
  {McConnachie}}]{Simard2011BulgeDiskDecompCatalog}
{Simard}, L., {Mendel}, J.~T., {Patton}, D.~R., {Ellison}, S.~L., \&
  {McConnachie}, A.~W. 2011, \apjs, 196, 11

\bibitem[{{Vibert} {et~al.}(2009){Vibert}, {Zamojski}, {Conseil}, {Llebaria},
  {Arnouts}, {Milliard}, \& {Guillaume}}]{Vibert2009SPIEEMphot}
{Vibert}, D., {Zamojski}, M., {Conseil}, S., {et~al.} 2009, in Society of
  Photo-Optical Instrumentation Engineers (SPIE) Conference Series, Vol. 7246,
  Computational Imaging VII, ed. C.~A. {Bouman}, E.~L. {Miller}, \&
  I.~{Pollak}, 72460U

\bibitem[{{Watkins} {et~al.}(2022){Watkins}, {Salo}, {Laurikainen},
  {D{\'\i}az-Garc{\'\i}a}, {Comer{\'o}n}, {Janz}, {Su}, {Buta}, {Athanassoula},
  {Bosma}, {Ho}, {Holwerda}, {Kim}, {Knapen}, {Laine},
  {Men{\'e}ndez-Delmestre}, {Peletier}, {Sheth}, \&
  {Zaritsky}}]{Watkins2022S4GExtended}
{Watkins}, A.~E., {Salo}, H., {Laurikainen}, E., {et~al.} 2022, \aap, 660, A69

\end{thebibliography}
\bibliographystyle{aasjournal}

\end{document}